\documentclass[11pt,a4paper]{emulateapj}
\usepackage{epsfig}
\usepackage{amsmath}

\usepackage{graphicx, amssymb} 
\usepackage{amsfonts}
\usepackage{bm} 

\usepackage[linktocpage]{hyperref}
\usepackage[caption=false]{subfig}
\usepackage[usenames]{color}

\def\be{\begin{equation}}
\def\ee{\end{equation}}

\newcommand{\bea}{\begin{eqnarray}}
\newcommand{\eea}{\end{eqnarray}}
\newcommand{\ben}{\begin{enumerate}}
\newcommand{\een}{\end{enumerate}}
\newcommand{\bi}{\begin{itemize}}
\newcommand{\ei}{\end{itemize}}

\newcommand{\nn}{\nonumber}

\def\ga{\mathrel{\raise.3ex\hbox{$>$\kern-.75em\lower1ex\hbox{$\sim$}}}}
\def\la{\mathrel{\raise.3ex\hbox{$<$\kern-.75em\lower1ex\hbox{$\sim$}}}}

\def\l{\left}
\def\r{\right}
\def\be{\begin{equation}}
\def\ee{\end{equation}}

\def\I_M{{I_{\scriptscriptstyle M\times M}}}

\def\be{\begin{equation}}
\def\ee{\end{equation}}
\def\bea{\begin{eqnarray}}
\def\eea{\end{eqnarray}}

\def\pa{\partial}

\begin{document}

\title{Into the lair: gravitational-wave signatures of dark matter}

\author{Caio F. B. Macedo\altaffilmark{1,2}, Paolo Pani\altaffilmark{2,3}, Vitor Cardoso\altaffilmark{1,2,4,5}, Lu\'is C. B. Crispino\altaffilmark{1}}

\affil{$^1$Faculdade de F\'{\i}sica, Universidade 
Federal do Par\'a, 66075-110, Bel\'em, Par\'a, Brazil} 
\affil{$^2$CENTRA, Departamento de F\'{\i}sica, 
Instituto Superior T\'ecnico,\\ Universidade T\'ecnica de Lisboa - UTL,
Avenida~Rovisco Pais 1, 1049 Lisboa, Portugal}
\affil{$^3$Institute for Theory $\&$ Computation, Harvard-Smithsonian
CfA, 60 Garden Street, Cambridge, MA, USA}
\affil{$^4$Perimeter Institute for Theoretical Physics
Waterloo, Ontario N2J 2W9, Canada}

\affil{$^5$Department of Physics and Astronomy, The University of Mississippi, University, MS 38677, USA.}


\begin{abstract}
The nature and properties of dark matter are both outstanding issues in physics. 
Besides clustering in halos, the universal character of gravity implies that self-gravitating compact dark matter configurations --~predicted by various models~-- might be spread throughout the universe.
Their astrophysical signature can be used to probe fundamental particle physics, or to test alternative descriptions of compact objects in active galactic nuclei. Here we discuss the most promising dissection tool of such configurations: the inspiral of a compact stellar-size object and consequent gravitational-wave emission. The inward motion of this ``test probe'' encodes unique information about the nature of the supermassive configuration.
When the probe travels through some compact region we show, within a Newtonian approximation, that the quasi-adiabatic inspiral is mainly driven by dark matter accretion and by dynamical friction, rather than by radiation-reaction. When accretion dominates, the frequency and amplitude of the gravitational-wave signal produced during the latest stages of the inspiral are nearly constant. 
In the exterior region we study a model in which the inspiral is driven by gravitational- and scalar-wave emission, described at fully relativistic level. Resonances in the energy flux appear whenever the orbital frequency matches the effective mass of the dark matter particle, corresponding to the excitation of the central object's quasinormal frequencies. Unexpectedly, these resonances can lead to large dephasing with respect to standard
inspiral templates, to such an extent as to prevent detection with matched filtering techniques.
We discuss some observational consequences of these effects for gravitational-wave detection.
\end{abstract}



\maketitle


\section{Introduction} 
The Universe is populated with a plethora of different gravitationally-bound objects in dynamical equilibrium. 
Luminous, hydrogen-fueled stars like our Sun are supported against collapse by radiation and gas pressure, whereas darker, compact and quiescent objects like neutron stars are prevented from full collapse by degeneracy pressure. Although evidence has been mounting for decades, 
only in recent years has it has become apparent that a completely different class of objects may, or {\it must}, also abound. Dark matter (DM) makes up a large fraction of galaxies and even though its exact nature is not known, it conglomerates into huge halos around the center of galaxies ~\citep{Bertone:2004pz}. 
Because all forms of matter gravitate, compact self-gravitating DM configurations could therefore be a substantial component of our own galaxy.

DM structures may form under a variety of scales and compactnesses. Typical DM halos are extended, low-density (by astrophysical standards) structures, but compact configurations are expected to form in several scenarios. One popular instance are compact boson stars~\citep{Liebling:2012fv}, which may form through collapse of scalar fields. The collapse can be triggered by a classical Jeans instability, gravitational cooling or even second-order phase transitions \cite{Bianchi:1989,Grasso:1990zg,Jetzer:1991jr,Seidel:1993zk,Frieman:1988ut}. It is also possible that compact DM profiles exist as extremely long-lived hair of rotating black holes, the endstate of superradiant instabilities \cite{Arvanitaki:2010sy,Pani:2012vp,Pani:2012bp,Witek:2012tr,Barranco:2012qs}.

Compact DM objects have also been occasionally invoked as an alternative to one of the most intriguing predictions of general relativity, namely the existence of black holes (BHs).  
Very massive main-sequence stars are dynamically unstable, a feature which is shared by most of the known compact configurations. 
Thus, ``standard'' stars cannot explain the dark, compact and supermassive objects lurking at the center of most galaxies, like the $\sim 10^6 M_{\odot}$ object in our own Milky Way~\citep{Genzel:1966zz}. It is widely accepted that BHs are the most natural explanation for these supermassive compact objects. Nevertheless, although actual observations support the BH hypothesis, experiments showing the direct existence of an event horizon are still missing. In fact, some argue that an observational proof of the event horizon based only on electromagnetic observation is fundamentally impossible~\citep{Abramowicz:2002vt}. In an attempt to test the BH paradigm, exotic forms of matter which possibly collapse to form supermassive horizonless objects, have been proposed. Besides their relevance for testing fundamental aspects of gravity, these objects may contribute to the dark matter content of the Universe~\citep{Bertone:2004pz}, being thus relevant for particle physics and cosmology. The exact nature of the object at the center of our galaxy will soon be strongly constrained by observations \citep{Doeleman:2008qh,Fish:2010wu,Eisenhauer:2008tg}, making this an exciting time to theoretically model and understand alternatives.

Common approaches to probe DM in astrophysical settings are based on model-dependent interactions between the DM and the baryonic sector. Such approaches usually focus on the imprints these interactions leave on the evolution and equilibrium structures of astrophysical objects. However, the equivalence principle guarantees that all forms of matter gravitate as predicted by Einstein's general relativity, regardless of the (non)baryonic nature of the constituent particles. As such, model-independent signatures of DM can arise from the study of gravitational self-interacting effects, like the existence of supermassive DM configurations. See~\citep{Eda:2013gg} for a very recent proposal in this direction.

It is widely believed that measurements of gravitational waves (GWs) from inspiralling stellar-mass objects into supermassive compact objects will map the entire spacetime geometry~\citep{Ryan:1995wh,Ryan:1997hg} and will carry imprints of the nature of the central object. Thus, GW measurements are in principle able to test the existence of compact DM objects and to discriminate between between BHs and other types of horizonless objects~\citep{Kesden:2004qx,Pani:2012zz}.

With the above as motivation, we study distinctive features of the extreme mass-ratio inspiral (EMRI) around supermassive scalar-field configurations. In this system, a small compact object with mass $\mu_p\in (1,10) M_{\odot}$ spirals into a supermassive object with mass $M \in (10^{4},10^{7}) M_{\odot}$. The late-time inspiral of this system  is of interest to future space-based GW detectors~\citep{AmaroSeoane:2012km,AmaroSeoane:2012je}. The typical orbital period is $(10^{2},10^{4})$ seconds and low-frequency GWs are emitted in the ($10^{-4},10^{-2}$)~Hz frequency band. In the EMRI limit, the inspiral can last tens to hundreds of years in the detector band. In one year of observation, millions of radians are contained in the signal, encoding rich information about the spacetime dynamics~\citep{Ryan:1995wh}. For this reason EMRIs are exceptional probes of strong-field gravity~\citep{Gair:2012nm}, modifications of general relativity~\citep{Cardoso:2011xi,Pani:2011xj,Yunes:2011aa,Yagi:2012vf} and of the nature of supermassive objects.

Here we argue that the GW signal from an EMRI can be also used to probe the existence of exotic fields that constitute the dark content of the Universe. If the central object is a BH, the inspiral terminates with a merger and subsequent ringdown~\citep{Berti:2009kk}. Instead, if the central object is formed by some compact DM configuration that interacts very weakly with baryonic stars, the EMRI proceeds also in the interior of the object, contributing significant amounts of signal-to-noise ratio to the signal~\citep{Kesden:2004qx}.
EMRIs are relatively clean systems that can be described with great accuracy within a perturbative approach. During most of the inspiral the stellar-mass object can be considered as a test particle moving on a fixed background. The timescale for merger is much longer than the orbital
period and the evolution can be described by an adiabatic approximation. At each instant, we consider that the particle follows a geodesic of the background spacetime and the secular evolution of the geodesic parameters can be computed by solving the linearized Einstein's equations. In this way one finds the inspiralling orbit and the corresponding gravitational waveform. This procedure takes into account the main dissipative effects of the back-reaction. A more detailed analysis, which would also consider conservative effects~\citep{Poisson:2003nc,Barack:2009ux} is beyond the scope of this work.

Among other models for self-gravitating exotic fields (e.g. axion stars~\citep{Kolb:1993zz}, boson-fermion stars~\citep{Henriques:1989ez}, etc...), boson stars (BSs) are particularly interesting because they arise as simple solutions of the Einstein-Klein-Gordon equations, without requiring any exotic matter other than a massive bosonic field (for reviews on the subject see~\citep{Jetzer:1991jr,Schunck:2003kk,Liebling:2012fv}). BSs are compact stars configurations that may be thought of as a natural realization of Wheeler's geons~\citep{Wheeler:1955zz} for scalar fields. Unlike the original geons, BSs can admit stable configurations which share many features with central galactic objects, without having singularities nor horizons~\citep{Torres:2000dw,Guzman:2005bs}, being indistinguishable from BHs in certain regimes. Formation of BSs has been studied extensively in the literature~\citep{Frieman:1988ut, Ferrell:1989kz, Frieman:1989bx, Grasso:1990zg, Madsen:1990gg, Tkachev:1991ka, Liddle:1993ha, Schunck:2003kk}. The recent discovery of the Higgs boson~\citep{Higgs:2012gk} is of course a further motivation to study this type of solutions. 

BSs can be classified~\citep{Schunck:2003kk} according to the scalar potential in the Klein-Gordon Lagrangian. 
Depending upon the scalar self-interactions, the maximum mass of a BS spans the entire range from one to billions of solar masses. 

Rather than working on a case-by-case analysis, here we focus on generic features that can leave a characteristic imprint on the waveform. The inspiral can be divided into two different regimes: the motion in the exterior of the supermassive object and the motion in the interior. 
In Section~\ref{sec:newtonian} we discuss EMRIs \emph{inside} generic DM configurations. We show that the inspiral is mostly driven by accretion of the scalar field onto the small compact object and by dynamical friction~\citep{Chandrasekhar:1943ys,Ostriker:1998fa}, rather than by GW emission. We include these effects in a Newtonian analysis and compute the signal emitted in GWs. The signal is markedly different from that arising during the merger into a supermassive BH and it also deviates substantially from the evolution obtained when accretion and gravitational drag are neglected~\citep{Kesden:2004qx}.
While our results of Section~\ref{sec:newtonian} are fairly generic, to discuss the outer evolution we need to specify some relativistic model. This is done in Section~\ref{sec:relativistic}, where we describe the EMRI around a spherically symmetric BS. We show that the evolution is driven by the emission of gravitational and scalar waves, which we describe at fully relativistic level. We show that, during a quasi-circular evolution, the energy flux can be resonantly excited. This leads to a large dephasing with important observational consequences for GW detection. 
We conclude in Section~\ref{sec:conclusions} by discussing possible extensions of our approach.
We use the signature $(-,+,+,+)$ for the metric and in most of the paper we adopt natural units $\hbar=c=G=1$, unless otherwise stated.

\section{GW-signatures of EMRIs inside compact DM configurations}\label{sec:newtonian}
We are interested in discussing simple models that can capture the salient features of the inspiral through a DM medium in the extreme mass-ratio limit.
There is a variety of situations in which such inspiral can occur. The most obvious one is the existence of self-gravitating, compact DM objects. Another is the inspiral around Kerr BHs surrounded by bosonic clouds~\citep{Arvanitaki:2010sy,Yoshino:2012kn,Pani:2012vp,Witek:2012tr}. Very recently, the GW signatures of intermediate mass BHs with DM mini-halos has also been considered~\citep{Eda:2013gg}.

Despite the multitude of models one can conceive, two of the most important generic features of these configurations are: 
\noindent (i)~ they interact with standard baryonic matter by purely gravitational effects and 
\noindent (ii)~they are typically supported by self-interactions of a massive bosonic field, whose mass can range between ${\cal O}(10^{-20})$~eV (or smaller) and ${\cal O}(1)$~TeV (or larger). In fact, as far as purely gravitational effects are considered, many dynamical aspects do not even depend on the nature of the DM particles (e.g. spin, mass, coupling constants,...), but solely on their mass-energy distribution.

Although compact DM configurations usually require relativistic effects to support their selfgravity, many features of the inspiral can be captured by a simple Newtonian model. In this section, we consider compact Newtonian objects which are characterized by some nonvanishing density profile $\rho(r)$ when $r<R$ and zero outside.
The simplest model is a constant-density star\footnote{Constant density solutions exist -- both in Newtonian gravity and in general relativity -- also when the fluid is anisotropic, as in the case of selfgravitating scalar fields~\citep{Dev:2000gt}. However, we will show that possible anisotropies in the fluid are not important to model the inspiral, because accretion and dragging are mostly insensitive to pressure in the extreme mass-ratio limit.} supported by purely radial pressure, but more realistic nonconstant profiles are also possible. 
The crucial point here is to assume that these objects do not couple with standard matter in any way other than through gravitational interactions. 
In particular, the small orbiting perturber can penetrate the stellar surface and move \emph{inside} the object. Our main interest here is to understand this motion and the corresponding GW emission.

\subsection{Accretion- and gravitational drag-driven inspiral}
While the motion in the exterior is driven by radiation-reaction only, when the particle penetrates the stellar surface other effects must be taken into account, namely the accretion of the nonbaryonic mass onto the small compact object and the drag force due to the gravitational interaction of the orbiting perturber with its own wake. In the context of EMRIs, these effects have been taken into account to study the imprints of matter surrounding supermassive BHs~\citep{Barausse:2007dy} on the GW emission. We will show that in the interior of compact nonbaryonic matter configurations, these processes are actually dominant and the inspiral is mostly driven by DM accretion rather than by GW emission.
\subsubsection{Accretion: Collisionless versus Bondi-Hoyle}
%
\begin{figure}
\begin{center}
\begin{tabular}{c}
\epsfig{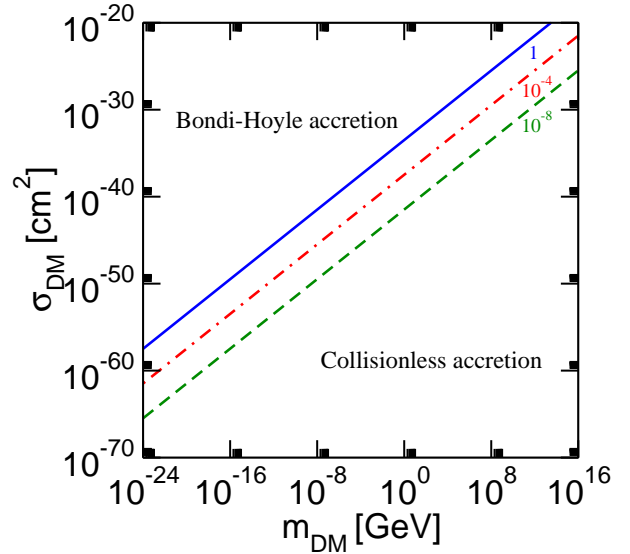}
\end{tabular}
\end{center}
\caption{\label{fig:meanfreepath}Level plots corresponding to the ratio $R_p/\ell=1,10^{-4},10^{-8}$ in the $(\sigma_{\rm DM},m_{\rm DM})$ plane, where $\sigma_{\rm DM}$ is the total self-interaction cross section for DM and $m_{\rm DM}$ is the mass of the DM particle. In the left-uppermost part of the parameter space the radius of the small compact object $R_p$ is much larger than the mean free path $\ell$ and accretion occurs at the Bondi-Hoyle rate~\eqref{Bondi}. In the lower-rightmost part of the diagram $R_p\ll\ell$ and accretion is governed by Eq.~\eqref{eqmu}. Straight lines refer to $\mu_p/M=10^{-6}$ and $R=2M$, but they have a simple scaling with the inverse of the mass ratio and with the compactness of the central object.
}
\end{figure}
Accretion of the scalar field onto the small compact object produces external forces that contribute to the secular evolution.
As long as the accreted mass is much smaller than the total mass of the orbiting object,
the assumption of quasi-stationary motion should provide a fairly accurate description.
Let us start by some simple estimates, assuming the accretion cross section is roughly the geometrical one of the small compact object.
For head-on collisions, the small compact object of mass $\mu_p$ traverses the entire diameter $2R$ of the star, accreting a tube of length $2R$ and 
radius $R_p$. Therefore, 
\be
\frac{M_{\rm accreted}}{\mu_p}\sim\frac{3}{2}\frac{\mu_p M}{R^2}\sim 0.02 \frac{\mu_p}{M}\ll1\,, 
\ee
where we used $R_p\approx\mu_p$, $R\sim 10 M$ and we have assumed constant density. Thus, during a single passage, the accreted matter has a negligible effect for head-on collisions. On the other hand, during the inspiral from the surface, the orbiting object can accrete much more as it sweeps through the equatorial plane, tearing a disk-gap of width $\mu_p$ and area $\sim R^2$. In this case we get
\be
\frac{M_{\rm accreted}}{\mu_p}\sim\frac{3M}{4R}\sim {\cal O}(1)\,,
\ee
for compact central objects. Thus, inspirals have to be carefully considered. The estimate above assumes that during the inspiral the probe does not travel through a region of depleted density caused by previous accretion, i.e. that the orbital timescale is longer than the replenishment timescale of the medium. The latter depends on the details of the model and on the nature of the DM medium. If the replenishment timescale is larger than the orbital period, the presence of gaps must be included, see e.g.~\citep{Lubow:1999kn,Lubow:2005ir,delValle:2012jq}. For simplicity, we neglect this possibility in the following analysis.

More rigorously, dynamical effects and the nature of the small compact object must be included. 
Accretion is described by
\be
\dot{\mu}_p=\sigma \rho v\,, \label{dmudt}
\ee
where $\rho$ is the density of the DM configuration, $v$ is the modulus of the velocity of the small object with respect to distant static observers and $\sigma$ is the accretion cross section. The latter strongly depends on the physical processes involved in the accretion and on the nature of the perturber. If the small compact object is a BH, whose radius $R_p$ is much smaller than the mean free path $\ell=(\sigma_{\rm DM} n)^{-1}$ ($\sigma_{\rm DM}$ and $n$ being the DM self-interaction total cross section and the particle density, respectively), then an approximate formula is $\sigma=\pi R_{\rm eff}^2$ where $R_{\rm eff}$ is the effective capture radius. For nonrelativistic free particles the latter reads~\citep{Unruh:1976fm,Shapiro:1983du,Giddings:2008gr}
\begin{equation}
 R_{\rm eff}\sim \frac{R_p}{v}\,. \label{Reff}
\end{equation}
This relation shows that, as $v\to0$, any particle that forms the supermassive object will eventually fall into the small BH even when orbiting at large distance. Therefore, the cross section in Eq.~\eqref{dmudt} is effectively infinite for accretion of particles at rest. Clearly, there exists a cutoff distance given by the radius $R$ of the supermassive star and which corresponds to a minimum velocity, $v_{\rm min}=R_p/R$. In the small mass-ratio limit $v_{\rm min}\sim \mu_p/M\ll1$. Therefore, for any $v>v_{\rm min}$ we get
\begin{equation}
 \dot{\mu}_p=\frac{\pi \rho R_p^2}{v} \qquad R_p\ll\ell\,. \label{eqmu}
\end{equation}
If the small compact object is a neutron star, the effective cross section is the minimum between the geometrical cross section and sum of the cross sections upon the individual nuclei of the star (see, e.g.,~\citep{Gould:1989gw,Bertone:2007ae}). Provided the scattering cross section between DM particles and the stellar nucleons is larger than $10^{-45}{\rm cm}^2$, Eq.~\eqref{eqmu} is still a good approximation for the accretion rate at nonrelativistic velocities. 
If the small object is a white dwarf, the geometrical cross section is typically larger than the individual scattering contributions, so that the effective cross section is given by the sum of the individual nuclei cross sections~\citep{Bertone:2007ae}. In this case the accretion rate will depend on the details of the microphysics and on the type of DM particles that make up the supermassive object. However, white dwarfs are not compact enough to sustain tidal forces and are likely to be tidally disrupted during the latest stage of the external inspiral. Thus, we shall restrict our attention to accretion onto relatively slow BHs and neutron stars only, both governed by Eq.~\eqref{eqmu}.

On the other hand, if the radius of the object is comparable to or larger than the mean free path, $R_p\gg\ell$, then accretion becomes a macroscopic process and cohesion forces and matter compressibility must be taken into account~\citep{Shapiro:1983du,Giddings:2008gr}. When the small compact object is a BH, this type of accretion is described by the Bondi-Hoyle formula~\citep{Bondi:1952ni,Bondi:1944jm,Shapiro:1983du}
\begin{equation}
 \dot{\mu}_p=4\pi\lambda \frac{\rho \mu_p^2}{(v^2+c_s^2)^{3/2}} \qquad R_p\gg\ell\,,\label{Bondi}
\end{equation}
where $c_s$ is the speed of sound and $\lambda$ is a number of order unity which depends on the details of the fluid. In our numerical simulations we have assumed $\lambda=1$, but the results depend on $\lambda$ very mildly. In Fig.~\ref{fig:meanfreepath}, we show the straight lines in the DM cross section-mass ($\sigma_{\rm DM}$--$m_{\rm DM}$) plane corresponding to $R_p/\ell=1,10^{-4},10^{-8}$ for a mass ratio $\mu_p/M_\odot=10^{-6} $ and $R=2M$. 
In the upper-leftmost part of the parameter space $R_p\gg\ell$ and accretion occurs at the Bondi-Hoyle rate~\eqref{Bondi}. In the lower-rightmost part of the diagram $R_p\ll\ell$ and accretion is governed by Eq.~\eqref{eqmu}. In the rest of this paper, we shall consider the two regimes separately.  Furthermore, following the nomenclature in~\cite{Shapiro:1983du}, we refer to the accretion governed by Eq.~\eqref{eqmu} as ``collisionless''. This only refers to the condition $R_p\ll\ell$ and not to the fact the the DM medium is pressureless (i.e. $\ell\to\infty$). In fact, collisionless accretion may also occur when the sound speed in the medium is nonvanishing --~and even possibly comparable with the speed of light~-- depending on the mass ratio and on the nature of the DM medium.
\subsubsection{Drag force}
Another important effect is the gravitational drag which results in a dynamical friction force on the small compact object traveling through the DM distribution~\citep{Chandrasekhar:1943ys,Ostriker:1998fa}. The gravitational field of the small perturber is felt universally, including by the DM
making up the compact configuration. Thus, a portion of this material is ``dragged'' along the inspiral, being tantamount to a net
decelerating force acting on the perturber. 
The theory of gravitational drag in collisionless media was developed in~\citep{Chandrasekhar:1943ys}. Although the self-interaction cross section for DM is typically small, compact objects can only be formed if the self-interaction is strong enough to support their own self-gravity, i.e. if the pressure is sufficiently high. This also means that the speed of sound in these compact objects is nonvanishing and typically of the same order of the speed of light. We discuss a particular example of such configuration in the next section. Therefore, for the class of objects we wish to describe, gravitational drag is appropriately described by dynamical friction~\citep{Ostriker:1998fa}.
Furthermore, at the scale of the small compact object in the EMRI limit, the density of the medium is nearly constant, so that we can adopt the theory of dynamical friction for motion through constant-density media. For linear motion, the dynamical force friction reads~\citep{Ostriker:1998fa}
\begin{equation}
 F_{\rm DF}=-\frac{4\pi  \mu_p^2\rho}{v^2}I_v\,, \label{ForceDF}
\end{equation}
with
\begin{eqnarray}
 I_v=\left\{\begin{array}{l}
           ×\frac{1}{2}\log\left(\frac{1+v/c_s}{1-v/c_s}\right)-v/c_s \,,\quad v<c_s\\
            \frac{1}{2}\log\left(1-\frac{c_s^2}{v^2}\right) +\log\left(\frac{vt}{r_{\rm min}}\right)\,,\quad v>c_s
          \end{array}\right. \label{IDF}
\end{eqnarray}
where $r_{\rm min}$ is the effective linear size surrounding the small compact object and, in case of supersonic motion, $r_{\rm min}(v-c_s)<t<R(v+c_s)$ is assumed~\citep{Ostriker:1998fa}.

The gravitational drag force in the case of circular motion has been derived in~\citep{Kim:2007zb}, where it was shown that the curvature of the orbit will bend the wake at large distances from the perturber. The subsonic motion is remarkably similar to the linear case, whereas the supersonic motion generically deviates from Eq.~\eqref{IDF}. In the latter case, the linear motion analysis is found to reproduce the exact results surprisingly well after the replacement $vt\to 2 r(t)$ in the last term on the second line of Eq.~\eqref{IDF}~\citep{Kim:2007zb}.

Note also that, within the extreme mass-ratio assumption, the motion effectively takes place on a linear trajectory at the scale of the small perturber. Consistently with this expectation, we have checked that the gravitational drag force obtained using Eq.~\eqref{ForceDF} [with the replacement $vt\to 2 r(t)$ in the supersonic regime] agrees with the fitting formulae given in~\citep{Kim:2007zb} for the gravitational drag force in circular motion. In what follows we adopt Eq.~\eqref{IDF} with the replacement $vt\to 2 r(t)$ to describe the drag force in circular motion.

It is also important to stress that very compact objects require high pressure to support their self-gravity. If the compactness is of the order of that of a neutron star or of a BH, the corresponding speed of sound $c_s$ might be comparable to the speed of light. Therefore, in the nonrelativistic limit we focus on, the motion is likely to be subsonic. Our primary goal is to discuss this subsonic regime but, for completeness, we shall also consider the case in which the orbital velocity exceeds the speed of sound. Supersonic motion is in principle allowed close to the surface and depending on the particular model at hand. Supersonic motion produces a sharp enhancement of the drag force when $v\approx c_s$~\citep{Ostriker:1998fa,Kim:2007zb}. In particular, the supersonic drag force depends on the parameter $r_{\rm min}$. In our models we have assumed $r_{\rm min}\equiv R_{\rm eff}\ll r$, and we have checked that different choices give qualitatively similar results. A detailed analysis of the supersonic case is beyond our scope and left for future work.

Dynamical friction might be very important during the inspiral. Indeed, since the force due to accretion reads $F_a=\dot\mu_p v$, we obtain
\begin{equation}\label{ratiomicro}
 \frac{|F_{\rm DF}|}{F_a^{\rm collisionless}}\sim \left\{\begin{array}{l} 
                                    \frac{1}{3}\frac{c^3}{c_s^3}\frac{v}{c}				\hspace{1.2cm}\quad v\ll c_s\\
                                    \log\left(\frac{2r(t)}{r_{\rm min}}\right)\frac{c^2}{v^2}		\quad v\gg c_s
                                   \end{array}\right.\,,
\end{equation}
for the accretion rate~\eqref{eqmu} and
\begin{equation}\label{ratioBondi}
 \frac{|F_{\rm DF}|}{F_a^{\rm Bondi}}\sim \left\{\begin{array}{l} 
                                    (3\lambda)^{-1}				\hspace{1.2cm}\quad v\ll c_s\\
                                    \lambda^{-1}\log\left(\frac{2r(t)}{r_{\rm min}}\right)		\quad v\gg c_s
                                   \end{array}\right.\,,
\end{equation}
for the Bondi rate~\eqref{Bondi}, respectively. In Eq.~\eqref{ratiomicro} we have considered $R_p\sim 2 \mu_p/c^2$ and we have reintroduced the speed of light $c$ for clarity. Thus, in the case of collisionless accretion, if the motion is supersonic but nonrelativistic, $c_s\ll v\ll c$, dynamical friction dominates over accretion. In the subsonic regime, if $c_s\sim {\cal O}(c)$, the drag force is subdominant in the nonrelativistic limit and we expect the late stages of the inner inspiral to be dominated by collisionless accretion effects. Nonetheless, depending upon the ratio $c_s/c$, there exists a crossover regime in which both effects are equally important. On the other hand, if accretion is governed by the Bondi rate, it is comparable to the gravitational drag force in the subsonic regime and it is negligible in the supersonic one. The results presented in the next section are obtained by including both the accretion and the drag force during the inspiral and considering both types of accretion separately.
\subsubsection{Gravitational radiation backreaction}\label{sec:reaction}
In addition to accretion and gravitational drag, the motion inside compact DM configurations is also driven by gravitational radiation reaction, similarly to the usual inspiral around compact objects in vacuum. Neglecting radiation-reaction, the motion of the perturber is governed by the gravitational force $F_g(r)=m(r)\mu_p/r^2$. For simplicity, in this paper we consider the following mass function and DM density profile: 
\begin{equation}
 m(r)=M\left(\frac{r}{R}\right)^\alpha\,, \qquad \rho(r)=\frac{\alpha  M}{4 \pi  R^{\alpha }}r^{\alpha -3}\,, \label{profile}
\end{equation}
but our results extend straightforwardly to more realistic profiles. Note that the equation above reduce to the case of constant-density DM configuration when $\alpha=3$ and to the vacuum case (briefly discussed in Appendix~\ref{app:Newtonian_ext}) when $\alpha=0$.

Neglecting radiation-reaction, the energy and angular momentum of the perturber are conserved quantities~\citep{Eda:2013gg}:
\begin{eqnarray}
 E&=&\frac{1}{2}\mu_p\dot r^2+\frac{L^2}{2\mu_p r^2}-\frac{M\mu_p}{(1-\alpha)R^\alpha r^{1-\alpha}}\,,\\
 L&=&\mu_p r^2\dot\varphi\,.
\end{eqnarray}
In the case of circular orbits of radius $r$, we get
\begin{eqnarray}
 E&=&\mu_p\frac{M(1+\alpha)}{R^\alpha r^{1-\alpha}2(\alpha-1)}\,, \label{EalphaR}\\
 L^2&=&\mu_p^2\frac{M}{R^\alpha}r^{1+\alpha}\,.
\end{eqnarray}
and the Keplerian frequency reads
\begin{equation}
 \Omega=\sqrt{\frac{m(r)}{r^3}}\,.\label{Omegat}
\end{equation} 
For circular orbits the radius and the orbital frequency are constant and, therefore, these orbits dissipate energy through the standard quadrupolar formula~\citep{Maggiore:1900zz}:
\begin{equation}
 \dot E_{\rm GW}=\frac{32 \mu_p ^2 M^3 r^{3\alpha-5}}{5 R^{3\alpha}}\,,
\end{equation}
where we have used Eq.~\eqref{Omegat}. Finally, differentiating Eq.~\eqref{EalphaR} and using the energy balance law, $\dot E=-\dot E_{\rm GW}$, we obtain the evolution equation of the orbital radius
\begin{equation}
 \dot r=-\frac{64 M^2\mu_p}{5(\alpha+1)r^{3-2\alpha}R^{2\alpha}}\,, \label{dotrreaction}
\end{equation}
whose solution reads
\begin{equation}
 r(t)=R\left(1+t/\tau_0\right)^{-\frac{1}{2(\alpha-2)}}\,, \label{Rreaction}
\end{equation}
with $(\alpha-2)\tau_0=5 (1 + \alpha)R^4/(128 M^2\mu_p)$. Note that the late time behavior of the solution above strongly depends on $\alpha$: for $\alpha<2$ the radius drops to zero in a finite time $|\tau_0|$, whereas for $\alpha>2$ the radius approaches zero asymptotically. In the singular case $\alpha=2$ the behavior is exponentially suppressed.

Let us now show that radiation-reaction effects are small compared to accretion. Simply by angular momentum conservation and using Eq.~\eqref{Omegat}, we obtain, for the secular evolution of the radius under accretion,
\begin{equation}
 \dot r_{\rm accretion}=-\frac{2}{1+\alpha}\frac{r(t)\dot\mu_p}{\mu_p(t)}\,. \label{draccretion}
\end{equation}
By comparing with Eq.~\eqref{dotrreaction}, we obtain
\begin{eqnarray}
 \frac{\dot{r}_{\rm collisionless}}{\dot{r}_{\rm reaction}}&\sim& \frac{5\alpha}{32}\left(\frac{R}{M}\right)^{3/2}\left(\frac{R}{r(t)}\right)^\frac{3(\alpha-1)}{2} \,,\\
 \frac{\dot{r}_{\rm Bondi}}{\dot{r}_{\rm reaction}}&\sim& \frac{5\alpha\lambda}{32 c_s^3}\frac{R}{M}\left(\frac{R}{r(t)}\right)^{\alpha-1} \,,
\end{eqnarray}
for collisionless and subsonic Bondi accretion, respectively. Interestingly, in both cases and for any $\alpha>1$, the late-time inspiral in the interior of DM compact configurations is generically dominated by accretion and not by radiative dissipation.

\subsection{Numerical evolution of the inspiral in the interior}
%
\begin{figure*}
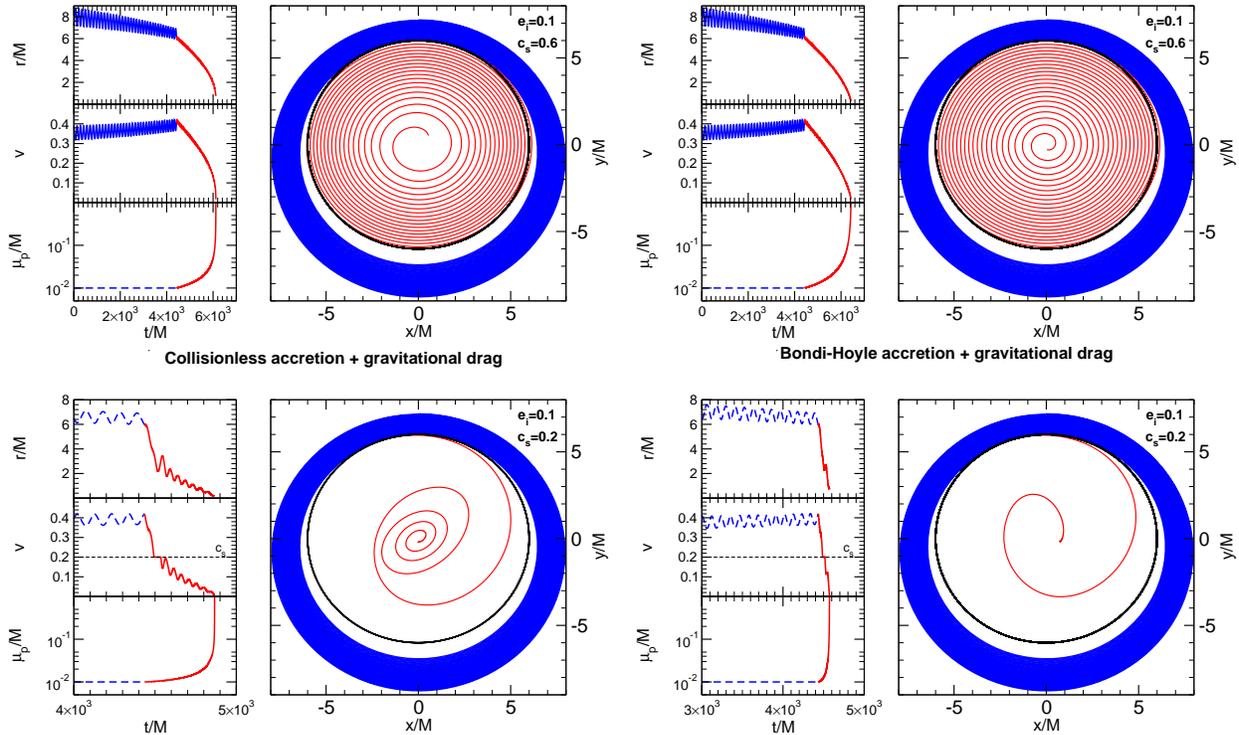

\begin{center}
\begin{tabular}{cc}
\epsfig{file=orbits_accretion.eps,width=8cm,angle=0,clip=true}&
\epsfig{file=orbits_accretion_Bondi.eps,width=8cm,angle=0,clip=true}\\
\epsfig{file=orbits_accretion_supersonic.eps,width=8cm,angle=0,clip=true}&
\epsfig{file=orbits_accretion_supersonic_Bondi.eps,width=8cm,angle=0,clip=true}
\end{tabular}
\end{center}
\caption{\label{fig:orbits_accretion}Secular evolution of the orbital parameters of a point particle orbiting a constant density, Newtonian star with radius $R=6M$. The particle starts at $r(0)=8M$ with initial eccentricity $e_i\equiv e(t=0)=0.1$. When $r>R$ (blue curves), the evolution is radiation-driven through the quadrupole formula (cf. Appendix~\ref{app:Newtonian_ext}). When $r<R$ (red curves) the evolution is driven by dynamical friction and by (i) collisionless accretion (left panels, cf. Eq.~\eqref{eqmu}) or ii) Bondi-Hoyle accretion (right panels, cf. Eq.~\eqref{Bondi}). Upper panels: $c_s=0.6$ and the motion is always subsonic. In the interior the orbits circularize quickly. Lower panels: $c_s=0.2$; the inspiral in the interior starts supersonic and the evolution is dominated by dynamical friction. When $v<c_s$, the evolution proceeds qualitatively as in the upper panel.
In the small left panels we show (from top to bottom): the radial position in polar coordinates, the module of the particle's velocity and the time evolution of the mass-ratio. The evolution does not qualitatively depend on the accretion rate formula used.
}
\end{figure*}
In this section we describe the secular evolution of the small perturber inside a spherically-symmetric DM configuration as driven by DM accretion and gravitational drag. As proved in the previous section, gravitational radiation reaction is a small effect compared to accretion, so we can safely neglect it here. 
In Newtonian theory, the accretion- and gravitational drag-driven inspiral is described by the following system:
\begin{eqnarray}
 \dot\mu_p \dot r+\mu_p(\ddot r-r\dot\varphi^2)&=&-\frac{\mu_p m(r)}{r^2}+F_{\rm DF,r} \,,\label{eqr}\\
 r \dot\mu_p \dot\varphi+\mu_p(r\ddot\varphi+2\dot r\dot \varphi)&=&F_{\rm DF,\varphi}\,, \label{eqtheta}
\end{eqnarray}
together with Eq.~\eqref{eqmu} or Eq.~\eqref{Bondi} for the evolution of $\mu_p(t)$. In the equations above $F_{\rm DF, r}$ and $F_{\rm DF, \varphi}$ are the two components of the gravitational drag force. Namely,
\begin{equation}
 F_{\rm DF, r}= F_{\rm DF}\frac{\dot r}{v}\,,\qquad F_{\rm DF, \varphi}= F_{\rm DF}\frac{r\dot \varphi}{v}\,.
\end{equation}
where $v^2=\dot r^2 +r^2\dot\varphi^2$ and $F_{\rm DF}$ is the linear dynamical force friction, Eq.~\eqref{ForceDF}. As discussed above, using the formula for linear motion with  the replacement $vt\to 2 r(t)$ is well justified in the extreme-mass ratio limit~\citep{Kim:2007zb}. 

Equation~\eqref{eqtheta} can be directly integrated in two extremal limits. Neglecting gravitational drag, and for \emph{any} accretion rate, we get
\begin{equation}
 r\Omega\equiv r\dot\varphi=\frac{L}{r \mu_p}\,,
\end{equation}
which can be also obtained from the conservation of the angular momentum $L$. On the other hand, if we neglect accretion $(\mu_p=$const) and in the limit $v\ll c_s$, we get
\begin{equation}
r\Omega\equiv r\dot\varphi=\frac{L}{r \mu_p}\exp\left(-\frac{4\pi\rho}{3 c_s^3}\mu_p t\right)\,,
\end{equation}
where we have assumed constant density.

In the general case, the system \eqref{eqr}-\eqref{eqtheta} has to be integrated numerically for a given density profile $\rho(r)$ and some initial conditions. The latter are chosen at the time the particle reaches the radius of the object at the end of the quasi-elliptical, radiation-driven inspiral in the exterior. 

Some results are shown in Fig.~\ref{fig:orbits_accretion}. As an example, we have considered a constant-density, Newtonian star with radius $R=6M$ and a point-like particle located at $r=8M$ at $t=0$. The initial eccentricity is $e_i=e(t=0)=0.1$. Similar results can be obtained for other choices of the parameters and for nonconstant density profiles\footnote{Nonetheless, in the deep interior of stellar configurations the density is nearly constant, so that we expect constant-density profiles to be a good approximation for the latest stages of the inner inspiral inside more complicate matter configurations.}. 

In absence of dissipative effects, the motion inside constant-density distributions is worked out in Appendix~\ref{app:spring_elipse}, where we show that
the small body moves on ellipses centered at the origin. The quasi-elliptical evolution is then governed by Eqs.~\eqref{eqr}-\eqref{eqtheta}. 
Figure~\ref{fig:orbits_accretion} summarizes the quasi-elliptical evolution of the orbital parameters in the subsonic (upper panels) and in the supersonic (lower panels) cases, both for collisionless accretion (left panels) and for Bondi accretion (right panels), respectively. In the top panels, we show the evolution of the orbital parameters for the case $c_s=0.6$ (in units of the speed of light), i.e. the motion in the interior of the object is always subsonic. In the large panels we show the orbit in cartesian coordinates. The small panels refer to the evolution of the orbital radius, the velocity and the mass of the small object, respectively. To help visualization, we have considered the unrealistic initial mass ratio $\mu_p(t=0)=10^{-2} M$ and the number of cycles grows with the inverse of $\mu_p/M$. In the interior, accretion is very efficient and the mass-ratio becomes of order unity during the inspiral, so that our approximation breaks down. 

In the interior (red curves) the orbit circularizes, regardless the details of the accretion rate. This can be proved analytically in the case of accretion-driven inspiral. Indeed, using the results of Appendix~\ref{app:spring_elipse}, together with conservation angular momentum conservation and the energy balance law, it is easy to derive the secular evolution of the semi-major axis $b$ and of the eccentricity in the small-eccentricity limit:
\begin{eqnarray}
 {\dot{b}_{\rm accretion}}&=&-b\left(\frac{1}{2}+\frac{e^2}{8}\right)\frac{\dot{\mu}_p}{\mu_p}+{\cal O}(e^4)\,. \label{dotb}\\
 {\dot{e}_{\rm accretion}}&=&-\frac{e}{4}\frac{\dot{\mu}_p}{\mu_p}+{\cal O}(e^3)\,. \label{dote}
\end{eqnarray}
The equation above shows the important result that circular motion remains circular. Because $\dot{\mu}_p>0$, orbits which are slightly eccentric
will tend to circularize. Therefore, as clearly shown in Fig.~\ref{fig:orbits_accretion}, circular inspiral is an attractor of the motion. Remarkably, this result is independent of the specific accretion process. 

In the bottom panels of Figure~\ref{fig:orbits_accretion}, we show the same quantities as in the top panels but for the case $c_s=0.2$, i.e. the motion in the interior starts supersonic and, as the velocity decreases, the inspiral enters the subsonic regime. As expected, during the supersonic phase the motion is dominated by dynamical friction and the velocity decreases abruptly. As the particle enters the subsonic regime, the evolution proceeds qualitatively as in Fig.~\ref{fig:orbits_accretion}. However, the inspiral in the supersonic case is much faster and the orbits do not have time to circularize. Note that the evolution does not qualitatively depend on the accretion rate formula used: the results shown in the left and right panels of Figure~\ref{fig:orbits_accretion} are qualitatively similar.

Finally, note that the accretion can be extremely efficient. In the latest stages of the inspiral, the orbiting object will start accreting very fast [cf. \eqref{mufinal} below] and an arbitrarily small initial mass will accrete an amount of matter $M$ in a finite amount of time. Therefore, our small mass-ratio hypothesis will eventually break down. If the orbiting object is a neutron star, during the inspiral it will develop a DM core and it will likely collapse to form a BH~\citep{Bertone:2007ae}. The signal emitted during the collapse is another distinctive feature that must be quantified by a relativistic analysis. 

%

\subsection{Gravitational waveforms}
\begin{figure*}
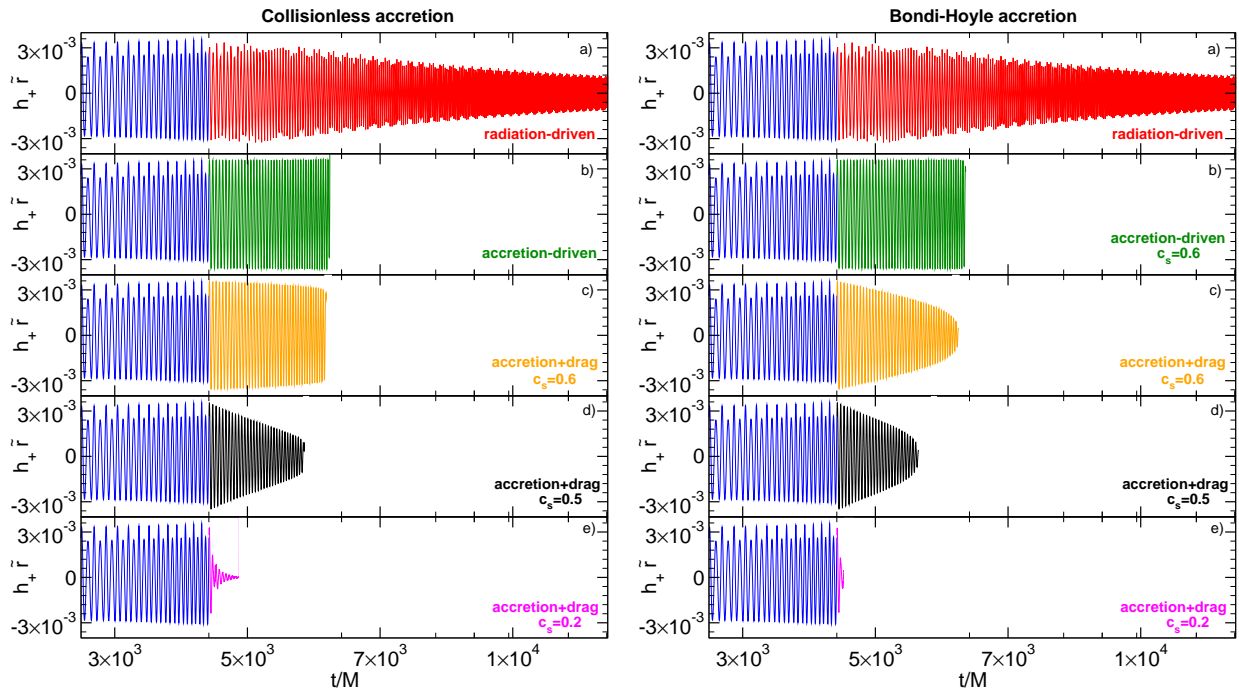

\begin{center}
\begin{tabular}{cc}
\epsfig{file=waveform_accretion.eps,width=8cm,angle=0,clip=true}&
\epsfig{file=waveform_accretion_Bondi.eps,width=8cm,angle=0,clip=true}
\end{tabular}
\end{center}
\caption{\label{fig:waveform_accretion}Gravitational-wave amplitude $h_+ \tilde{r}$ along $(\iota,0)=(\pi/2,0)$ as a function of time when collisionless accretion (left panels) and Bondi accretion (right panels) are considered, respectively. Small panels refer to: a) usual radiation-driven inspiral (neglecting accretion and dynamical friction); b) accretion-driven inspiral (neglecting radiation and dynamical friction); c) inspiral driven by accretion and dynamical friction (neglecting radiation) with $c_s=0.6$, which corresponds to the orbits shown in the top panels of Fig.~\ref{fig:orbits_accretion}; d) same as panel c) but with $c_s=0.5$; e) supersonic regime with $c_s=0.2$, which corresponds to the orbits shown in the bottom panels of Fig.~\ref{fig:orbits_accretion}. Remaining parameters as in Fig.~\ref{fig:orbits_accretion}.
}
\end{figure*}
Here we present the waveforms emitted during the inspiral shown in Fig.~\ref{fig:orbits_accretion}. Once the orbital radius, angular velocity and perturber mass are obtained as functions of time, we can use the standard quadrupole formula to compute $h_+(t)$ and $h_\times(t)$ for a distant observer located at $\tilde{r}$ with an angle view of $(\iota,0)$ [cf. Eq.~(3.72) in ~\citep{Maggiore:1900zz}].

In Figure~\ref{fig:waveform_accretion} we consider $\iota=\pi/2$, so that only $h_+$ is nonvanishing and the wave is linearly polarized.
The motion in the exterior is driven by the classical radiation-reaction mechanism, briefly summarized in Appendix~\ref{app:Newtonian_ext} for completeness.
In panel a) we have neglected accretion and dynamical friction, so that the evolution proceeds only through the radiation-backreaction. In panel b) we have considered accretion but neglected gravitational drag and the signal has a constant amplitude and constant frequency. We discuss this case in detail in the next section. In panels c)-e) we have included both accretion and dynamical friction with different constant values of $c_s$. 

The last panel shows the supersonic case described in the lower set of panels of Figure~\ref{fig:orbits_accretion}. As expected, the contribution of dynamical friction becomes dominant as $c_s\ll c$ and we observe two effects: (i) the amplitude of the signal decreases in time and (ii) the total time of the inspiral in the interior quickly decreases for smaller values of $c_s$. The latter effect is simply due to the extra dissipative channel during the evolution. 

These features would be completely missed by an evolution purely driven by radiative effects. Our analysis does not include relativistic effects, but provides a strong case for including accretion and gravitational drag effects in the inspiral inside compact DM configurations.

\subsection{Analytical Fourier waveforms in the stationary-phase approximation}
Some of the qualitative features of the waveforms presented in Figure~\ref{fig:waveform_accretion} can be analytically captured by a toy model in which we consider a small object of mass $\mu_p$ on a quasi-circular orbit inside a Newtonian star and whose secular evolution is driven by a single dissipative effect.
At Newtonian order, it is possible to obtain analytical templates of the waveforms in Fourier domain in the case of accretion-driven and radiation-driven inspiral. To lowest order, $\mu_p$, $r$ and $\Omega$ are constant and the Newtonian waveforms simply read:
\begin{eqnarray}
 h_+(t)&=&\frac{G r^2 \mu  \omega_{\rm GW}^2 }{c^4 \tilde{r}}\left(\frac{1+\cos^2\iota}{2}\right)\cos( \omega_{\rm GW}t) \,, \label{hp}\\
 h_\times(t)&=&\frac{G r^2 \mu  \omega_{\rm GW}^2 }{c^4  \tilde{r}}\cos\iota \sin( \omega_{\rm GW}t)\,, \label{hm}
\end{eqnarray}
where $\omega_{\rm GW}=2\Omega$. Then, dissipative effects can be included by replacing the constant parameters $\omega_{\rm GW}$, $r$ and $\mu_p$ by $\omega_{\rm GW}(t)$, $r(t)$ and $\mu_p(t)$, where the secular time evolution is governed by the specific dissipative mechanism~\citep{Maggiore:1900zz}. In the next sections, we shall treat collisionless accretion, Bondi accretion and radiation-reaction separately.
\subsubsection{Collisionless accretion, $R_p\ll \ell$}
In order to isolate the effects of accretion, let us neglect GW reaction and gravitational drag. As we discussed above, if $c_s={\cal O}(c)$, the gravitational drag is a small effect compared to accretion, at least in the nonrelativistic regime and for collisionless accretion. Let us then consider the orbital evolution driven by accretion only. 
Because the system is in isolation, the total angular momentum
is constant, even when accretion is included. On the other hand, the binding energy evolves during accretion. 
A simple and powerful result can be obtained solely by angular momentum conservation,
\be
\mu_p(t) r(t)^2=\frac{L}{\Omega(t)}\,.\label{angmomentum_cons}
\ee
For quasi-circular orbits, $v=r\Omega$ and the orbital frequency reads as in Eq.~\eqref{Omegat}.
Using Eq.~\eqref{angmomentum_cons}, we obtain
\begin{equation}
 \mu(t)=\frac{L}{\sqrt{G m(r) r(t)}}\,, \label{mut}
\end{equation}
where, here and in the rest of the section, we restore factors $G$ and $c$ for clarity.
Finally, using Eq.~\eqref{eqmu} we get:
\begin{equation}
\dot r=-\frac{8\pi L G}{c^2}\frac{\rho(r) r(t)}{m(r)+4 \pi \rho(r) r(t)^3}\qquad R_p\ll \ell\,.\label{rdotmicro}
\end{equation}
Once a density profile is specified, the equation above can be solved for $r(t)$. Then, the other dynamical quantities $\mu(t)$ and $\Omega(t)$ respectively read as in Eqs.~\eqref{mut} and \eqref{Omegat}.

To be concrete, let us consider the density profile given in Eq.~\eqref{profile}.
In this case, Eq.~\eqref{rdotmicro} can be directly integrated:
\begin{eqnarray}
 r(t)&=&R\left(1-{t}/{t_{\rm insp}}\right)^\frac{1}{3} \,, \label{rt_collisionless} \\
 \mu_p(t)&=&\mu_p^{(i)}\left(1-{t}/{t_{\rm insp}}\right)^{-\frac{\alpha+1}{6}} \label{mufinal}\,, \\
 \Omega&=&\Omega_i \left(1-{t}/{t_{\rm insp}}\right)^\frac{\alpha-3}{6}\,,
\end{eqnarray}
where $\mu_p^{(i)}$ is the mass of the particle at the time $t=0$, with $R=r(0)$ and $\Omega_i=\Omega(0)$, and we have introduced the duration of the inspiral,
\be
t_{\rm insp}=\frac{c^2 R^3 (1+\alpha)}{6 G\alpha L}= \frac{c^2(1+\alpha)}{8\pi G^{3/2} \alpha\langle\rho\rangle \mu_p^{(i)}}\sqrt{\frac{M}{R}}\,,
\ee
where $\langle\rho\rangle=3M/(4\pi R^3)$.

With this solution at hand, we can compute the corresponding gravitational waveforms~\citep{Maggiore:1900zz}.
Standard treatment allows to write Eqs.~\eqref{hp} and \eqref{hm} to lowest order as
\begin{eqnarray}
 h_+(t)&=&A_+(t_{\rm ret})\cos\varpi(t_{\rm ret}) \,,\\
 h_\times(t)&=&A_\times(t_{\rm ret}) \sin\varpi(t_{\rm ret})\,,
\end{eqnarray}
where $t_{\rm ret}$ is the retarded time and $\varpi=2\int_{t_i}^t dt'\Omega(t')$. In the case at hand, we get
\begin{equation}
 \varpi=\varpi_0-\frac{12 t_{\rm insp} }{3+\alpha } \sqrt{\frac{GM}{R^3}}\left(\frac{\tau }{t_{\rm insp}}\right)^{\frac{3+\alpha }{6}}\,,
\end{equation}
where $\tau=t_{\rm insp}-t$ and $\varpi_0=\varpi(t_{\rm insp})$. Note that the orbital frequency $\Omega$ is constant when the density is homogeneous. 
When the density is nonconstant it is relevant to compute the Fourier transform of the waveform. Adopting a stationary phase approximation~\citep{Maggiore:1900zz}, we obtain
\begin{equation}
 \tilde{h}_+= {\cal A}_+e^{i\Psi_+}\,,\qquad  \tilde{h}_\times={\cal A}_\times e^{i\Psi_\times}\,,
\end{equation}
where 
\begin{eqnarray}
  {\cal A}_+&=&\frac{4\sqrt{3}i \mu_p^{(i)}}{c^4\tilde{r} }\sqrt{\frac{t_{\rm insp}}{2(\alpha -3)}}\frac{1+\cos^2\iota}{2}\nn\\
  &&\times\left[G^\frac{3(\alpha-4)}{2} \pi ^\alpha M^{\frac{\alpha-6 }{2}} R^{\frac{6+\alpha }{2}} f^{\frac{3+\alpha }{2}} \right]^\frac{1}{\alpha-3}\,,\\
 {\cal A}_\times&=& \frac{2\cos\iota}{1+\cos^2\iota}{\cal A}_+\,, \\
 \Psi_+&=& \frac{12t_{\rm insp} }{3+\alpha } \left(\frac{R^3}{G M}\right)^{\frac{3}{\alpha -3}}(\pi f)^{\frac{3+\alpha }{\alpha -3}}\nn\\
 &&+2\pi f\left(t_c+\frac{r}{c}\right)-\varpi_0-\frac{\pi}{4}\,,\\
 \Psi_\times&=&\Psi_+ + \pi/2\,.
\end{eqnarray}
where $f=2\Omega/(2\pi)$.
Therefore, for $\alpha>0$ the amplitude and the phase increase with the frequency. 

\subsubsection{Bondi-Hoyle accretion, $R_p\gg \ell$}
In this case, from Eq.~\eqref{Bondi} we get
\begin{equation}
\dot r= -\frac{8\pi L G\lambda}{ c_s^3}\frac{\rho(r)\sqrt{G m(r) r(r)}}{m(r)+4 \pi \rho(r) r(t)^3} \qquad R_p\gg\ell\,,\label{rdotBondi}
\end{equation}
which, for the density profile given in Eq.~\eqref{profile}, can be directly integrated
\begin{eqnarray}
 r(t)&=&R\left(1-{t}/{t_{\rm insp}^{(B)}}\right)^\frac{2}{7-\alpha} \,, \\
 \mu_p(t)&=&\mu_p^{(i)}\left(1-{t}/{t_{\rm insp}^{(B)}}\right)^{-\frac{\alpha+1}{7-\alpha}} \label{mufinalBondi}\,, \\
 \Omega&=&\Omega_i \left(1-{t}/{t_{\rm insp}^{(B)}}\right)^\frac{\alpha-3}{7-\alpha}\,,
\end{eqnarray}
where we have assumed $\alpha\neq7$ and
\be
t_{\rm insp}^{(B)}=\frac{c_s^3 M R^{7/2} (1+\alpha)}{(7-\alpha)\alpha\lambda L (GM)^{3/2}}= \frac{c_s^3(1+\alpha)}{4\pi G^2 \alpha(7-\alpha)\langle\rho\rangle \lambda \mu_p^{(i)}}\,.\nn
\ee
Similarly to the previous case, using the solution above we can perform a stationary phase approximation to compute the amplitude and the phase in the frequency domain. The relevant ones read
\begin{eqnarray}
  {\cal A}_+&=&\frac{2 i\mu_p^{(i)}}{c^4\tilde{r} } \sqrt{\frac{t_{\rm insp}^{(B)}(7-\alpha)}{\alpha -3}} \frac{1+\cos^2\iota}{2}\nn\\
  &&\times\left[G^\frac{7\alpha-25}{4}\pi ^\frac{1+\alpha}{2}  M^\frac{3\alpha-13}{4} R^\frac{15-\alpha}{4} f^{2}\right]^{1/(\alpha-3)}\,,\\
 \Psi_+&=& \frac{7-\alpha}{2}t_{\rm insp}^{(B)}\left(\frac{R^3}{GM}\right)^\frac{7-\alpha}{2(\alpha-3)}(\pi f)^\frac{4}{\alpha-3}\nn\\
 &&+2\pi f\left(t_c+\frac{r}{c}\right)-\varpi_0-\frac{\pi}{4}\,.
\end{eqnarray}

\subsubsection{Gravitational radiation-reaction}
The motion driven solely by gravitational radiation reaction has been worked out in Section~\ref{sec:reaction}. Using Eqs.~\eqref{Rreaction} and ~\eqref{Omegat} we obtain, to lowest order,
\begin{eqnarray}
  {\cal A}_+&=&\frac{4 i\mu_p^{(i)}}{c^4\tilde{r} } \sqrt{\frac{\tau_0(\alpha-2)}{\alpha -3}} \frac{1+\cos^2\iota}{2}\nn\\
  &&\times\left[G^{2\alpha-7}\pi^2  M^{\alpha-4} R^{6-\alpha} f^{\frac{7-\alpha}{2}}\right]^{1/(\alpha-3)}\,,\\
 \Psi_+&=& \frac{8(\alpha-2)}{5-3\alpha}\tau_0\left(\frac{GM}{R^3}\right)^\frac{2(\alpha-2)}{\alpha-3}\left(\pi f\right)^\frac{5-3\alpha}{\alpha-3}\nn\\
 &&+2\pi f\left(t_c+\frac{r}{c}\right)-\varpi_0-\frac{\pi}{4}\,.
\end{eqnarray}
where $\tau_0$ is defined below Eq.~\eqref{Rreaction}.

Comparing the waveform amplitudes and phase in Fourier space shows that each specific dissipative mechanism produces a very peculiar signal. In particular, the signal produced by accretion-driven inspiral
is dramatically different from that arising in the classical radiation-driven inspiral, already at Newtonian level. 
Our analysis can be easily extended to more realistic density profiles. In general, the signal strongly depends on the DM profile and the waveform parameters are also sensitive to the type of accretion process. Thus, the GW signal can be used to constrain the nature and the properties of compact DM configurations.

\section{Relativistic analysis for the external inspiral}\label{sec:relativistic}
A self-consistent relativistic analysis of the inspiral around scalar-field compact configuration is fairly involved already at first order in the mass ratio.
Indeed, due to the coupling between gravitational and scalar perturbations and to the presence of a background scalar field, a small perturber will also source scalar waves, even if it is formed by purely baryonic mass. On the other hand, this property allows for a much richer phenomenology that would be missed if a fully relativistic analysis is not properly performed. To be concrete, in this section we focus on specific models for selfgravitating DM objects.
We investigate the emission of gravitational and scalar waves sourced by a test-particle in circular orbit around a boson star (BS). In this section we discuss the main features of this process. The mathematical procedure is standard but technically involved. A detailed analysis will appear elsewhere~\citep{elsewhere}.

\subsection{Relativistic models of supermassive DM objects}\label{sec:models}
\begin{table*}
 \begin{tabular}{c | c | c | c | c | c | c | c }
  \hline\hline
& $V_s(|\Phi|^2)$ & $\phi_0(r=0)$  & $\omega/\mu$ & $M \mu$ & $R\mu$ & $M\omega $  & $R/M$ \\
 \hline  \hline
mini-BS I 	&$\mu^2|\Phi|^2$				&0.0541 & 0.853087 & 0.63300 & 7.86149 & 0.54000 &  12.4194 \\
mini-BS II 	&						& 0.1157 & 0.773453 & 0.53421 & 4.52825 & 0.41319 & 9.03368 \\\hline
massive-BS I	&$\mu^2|\Phi|^2+\frac{\lambda}{2}|\Phi|^4$	& 0.0188 & 0.82629992558783 &  2.25721 & 15.6565 &    1.86513 &  6.9362 \\
massive-BS II 	&						&0.0309 & 0.79545061700675 &  1.92839 & 11.3739 &    1.53394&   5.8981 \\\hline
solitonic BS I 	&$\mu^2|\Phi|^2 (1-2|\Phi|^2/\sigma_0^2)^2$	&0.0371 & 0.1220326382068426831501347644 & 7.36961 & 22.8587 &0.89933 & 3.1017\\
solitonic BS II &						&0.0389 & 0.1236174876926880171453994576& 6.77654 & 20.2924 & 0.83770 & 2.9945 \\
\hline\hline
\end{tabular}
\caption{Boson-star models used in this work. The quantities $M$, $R$ and $\mu$ represent the mass of the solution, its effective radius [such that the mass function at $r=R$ corresponds to $99\%$ of the total mass $M$, see e.g. Ref.~\cite{Schunck:2003kk} for a discussion of other definitions] and the mass of the scalar field. 
For massive-BS and solitonic-BS configurations we used the fiducial values $\lambda=800\pi\mu^2$ and $\sigma_0=0.05$. The labels ``I'' and ``II'' respectively refer to stable and unstable configurations with respect to radial perturbations. The significant digits of the eigenvalue $\omega$ do not represent the numerical precision, but they show the fine tuning needed to obtain the solution.}
\label{tableconf}
\end{table*}

For concreteness, in the following we focus on BS configurations~\citep{Liebling:2012fv} which are relativistic solutions of the Einstein-Klein-Gordon theory but most of our results hold at a qualitative level for different models.

We consider the Einstein-Klein-Gordon theory:
\be
S=\int d^4 x \sqrt{-g} \left[\frac{R}{2\kappa} -g^{ab}\partial_a\Phi^*\partial_b\Phi-V_s(|\Phi|^2)\right] +S_{\rm matter},\nn
\ee
where $\kappa=8 \pi$ and $S_{\rm matter}$ denotes the action of any baryonic matter field. The Einstein-Klein-Gordon equations read
\begin{eqnarray}
R_{ab}-\frac{1}{2}g_{ab}R &=& \kappa\l( T_{ab}^\Phi + T_{ab}^{\rm matter} \r)\,,\label{eineq}\\
\frac{1}{\sqrt{-g}}\pa_a \l(\sqrt{-g}g^{ab}\pa_b\Phi\r)&=&\frac{d V_s}{d|\Phi|^2}\Phi\,, \label{KGeq}
\end{eqnarray}
where
\be
T^\Phi_{ab}=\partial_a\Phi^*\partial_b\Phi+\partial_b\Phi^*\partial_a\Phi-g_{ab}\l(\partial^c\Phi^*\partial_c\Phi+V_s(|\Phi|^2)\r)\,,\nn
\ee
is the energy-momentum of the scalar field. For a complex scalar field, Eq.~\eqref{KGeq} is supplied by its complex conjugate.
We will focus on spherically symmetric selfgravitating objects, whose line element is
\be
ds_0^2=-e^{v(r)}dt^2+e^{u(r)}dr^2+r^2 (d\theta^2+\sin^2\theta d\varphi^2)\,, \label{metric0}
\ee
whereas the background scalar field reads
\be
\Phi_0(t,r)\equiv \phi_0(r)e^{-i\omega t}\,,
\ee
with $\phi_0(r)$ being a real function. 
The field equations for $v(r)$, $u(r)$ and $\phi_0(r)$ form an eigenvalue problem for $\omega$ and they can be solved by standard methods, see e.g.~\citep{Kaup:1968zz,Ruffini:1969qy,Colpi:1986ye,Gleiser:1988rq,Gleiser:1988ih,Schunck:2003kk}. The equations can be recast in the form of Einstein gravity coupled to an anisotropic, nonbarotropic fluid, whose energy density and pressure are defined in terms of the stress-energy tensor of the scalar field, $T^\Phi_{ab}$. Namely,
\bea
\rho&\equiv&-{{T^\Phi}_t}^t=\omega^2e^{-v}\phi_0^2+e^{-u}(\phi_0')^2+V_s^0\,,\\
p_r&\equiv&{{T^\Phi}_r}^r=\omega^2e^{-v}\phi_0^2+e^{-u}(\phi_0')^2-V_s^0\,,\\
p_t&\equiv&{{T^\Phi}_\theta}^\theta=\omega^2e^{-v}\phi_0^2-e^{-u}(\phi_0')^2- V_s^0\,.
\eea
where $V_s^0=V_s(\phi_0)$ and $\rho$, $p_r$ and $p_t$ are the density, radial pressure and tangential pressure of the fluid, respectively. Using a standard shooting method, we have constructed spherically symmetric compact BSs which are solutions of three different models, presented in Table~\ref{tableconf}. 

\subsection{Geodesics around boson stars \label{sec:geodesics}}
%
\begin{figure*}
\begin{center}
\begin{tabular}{c}
\epsfig{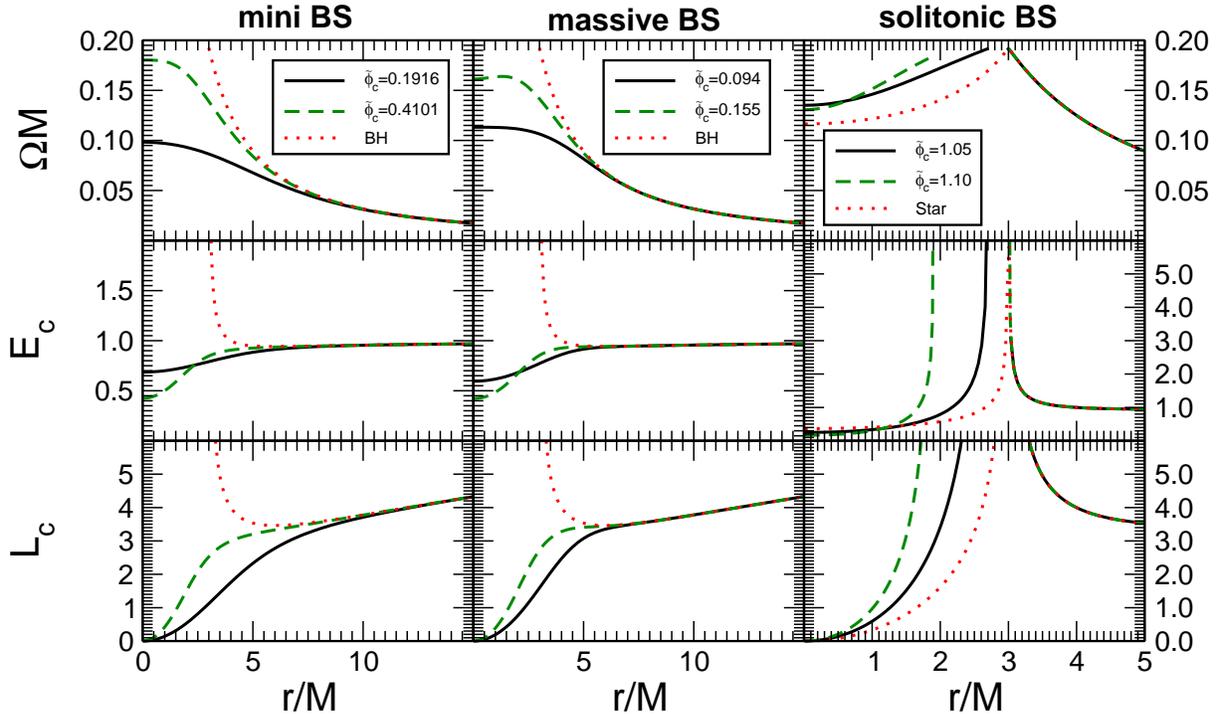}
\end{tabular}
\end{center}
\caption{\label{fig:circular_geodesics}
Circular geodesic motion for different BS models and configurations (cf. Table~\ref{tableconf}). In the top, middle and lower row we show the orbital frequency $\Omega$, the energy $E_c$ and the specific angular momentum $L_c$, respectively. Each column refers to a different BS model. From left to right: mini-BS, massive-BS and solitonic BS. For each model, we compare the geodesic quantities to those of a Schwarzschild BH and for the solitonic BS model we also compare to the metric elements of a uniform density star with $R=3M$. In the last column, the markers indicate the outer last stable orbit for solitonic BS configurations, which is approximately given by $r\approx 6M$ and $M\Omega_{isco}\approx 0.06804$. The light-rings are given by $r_{l-}\approx2.72093M$ and $r_{m}\approx2.9812M$, with $M\Omega_{l-} \approx 0.188818 $ and $M\Omega_{l+}\approx 0.192453$, for the first configuration and $r_{l-}\approx 1.91163M$ and  $r_{m}\approx2.99883M$, with $M\Omega_{l-} \approx 0.184590 $ and $M\Omega_{l+}\approx 0.192452$, for the second one.
}
\end{figure*}
Stellar-size objects gravitating around supermassive BSs have a small back-reaction on the geometry
and to first order in the object's mass move along geodesics of the BS background.
Accordingly, GW emission by such binaries requires a knowledge of geodesic
motion together with the consequent perturbative expansion of the gravitational field.
Many features of the gravitational radiation can be understood from the geodesic motion, in which we now focus.
We will also focus exclusively on circular, geodesic motion. The rationale behind this is that it makes the calculations much simpler
while retaining the main features of the physics. Furthermore, it can be shown that generic eccentric orbits get circularized by GW emission in vacuum \citep{Peters:1964zz}, on a time scale that depends on the mass ratio. 

We follow the analysis by~\citep{Chandrasekhar:1985kt}, the formalism for a generic background is presented in \citep{Cardoso:2008bp}. Following previous studies, we assume that the point-particle is not directly coupled to the background scalar field
\citep{Kesden:2004qx,Guzman:2005bs,Torres:2002td}. The conserved energy $E$, the angular momentum parameter per unit rest mass $L$, and the orbital frequency of circular geodesics read
\bea
E_c&=&\l(\frac{2 A(r_c)^2}{{2 A(r_c)-r_c A'(r_c)}}\r)^{1/2}\\
L_c&=&\l(\frac{r_c^{3} {A'(r_c)}}{2 A(r_c)-r_c A'(r_c)}\r)^{1/2},\\
\Omega&=&\frac{\dot{\varphi}}{\dot{t}}=\l(\frac{A'(r_c)}{2 r_c}\r)^{1/2}.
\eea
with $A(r)=e^v$ as defined in Eq.~\eqref{metric0}. A summary of the geodesic quantities corresponding to the BS models in Table~\ref{tableconf} is presented in Fig.~\ref{fig:circular_geodesics}.
Up to the innermost stable circular orbit of a Schwarzschild spacetime, $r=6M$, geodesic quantities are very close to their Schwarzschild counterpart with same total mass, as might be expected for very compact configurations. However, the geodesic structure at $r<6M$  can be very different~\citep{Kesden:2004qx}. A striking difference is that stable, circular timelike geodesics exist for BSs even well deep into the star~\citep{Torres:2002td,Guzman:2005bs,Kesden:2004qx}.

For less compact configurations -- namely mini BSs and massive BSs -- stable circular geodesics exist \emph{all the way down} the center of the star. This is an important feature because, if radiative effects and external forces are small, the inspiral will proceed through a secular evolution of these orbits. Furthermore, the orbital frequency $\Omega$ is roughly constant close to the origin. This also implies that the circular geodesics  deep inside the BS are nonrelativistic, as the velocity measured by static observers at infinity vanishes as the radius approaches zero. In this case, our previous Newtonian analysis should provide reliable results. Therefore, we expect that even the circular orbits in the interior of the object will be accessible to a quasi-circular evolution that, as we previously discussed, is mainly driven by accretion and gravitational drag effects. 

More compact configurations such as solitonic BSs may show truly relativistic effects. For these models $R\sim 3 M$ and an innermost stable circular orbit exists at $r\approx 6M$ with $M\Omega_{isco}\approx 0.06804$. This is to be expected, as the background scalar field is exponentially suppressed and the spacetime is very close to Schwarzschild outside the solitonic-BS effective radius. Like in the case of a Schwarzschild BH, there exists an unstable light ring at roughly $r_{m}\approx3M$.
The unexpected feature is the presence of a second {\it stable} light ring at $r_{l-}$, together with a family of stable timelike circular geodesics all the way to the center of the star. This is clearly a relativistic feature, which is similar to the case of uniform density stars. The latter may also present two light-ring and stable circular time-like orbits in their interior, depending on their compactness.
The bottom panels of Fig.~\ref{fig:circular_geodesics} depicts a uniform density star with radius $R=3M$. In this case, the two light-rings degenerate in the star surface. 
It has been argued~\citep{Kesden:2004qx} that, for these very compact models, the orbiting particle plunges when it reaches the innermost stable circular orbit and, due to radiation effects, the eccentricity of the orbits in the interior of the BS will increase. However, as we have discuss, accretion effects are dominant in the interior of the star and they also contribute to circularize the orbit. Our results are based on a Newtonian analysis, which is nonetheless accurate well deep inside the star, where the velocity is small. It is therefore possible that, after the initial plunge, the particle will have access to these circular geodesics, whose evolution is governed by accretion, rather than by radiative effects. 

\subsection{Point-particle orbiting a boson star, resonant fluxes and quasinormal modes \label{sec:PP}}
The gravitational and scalar energy fluxes emitted during the quasi-circular inspiral of a test-particle around a compact BS can be derived at fully relativistic level. The emission is governed by an inhomogeneous system of equations, whose regular solutions can be constructed via standard Green's function techniques. We present here the main results, whereas a detailed analysis will appear elsewhere~\citep{elsewhere}.

\begin{figure}
\begin{center}
\begin{tabular}{c}
\epsfig{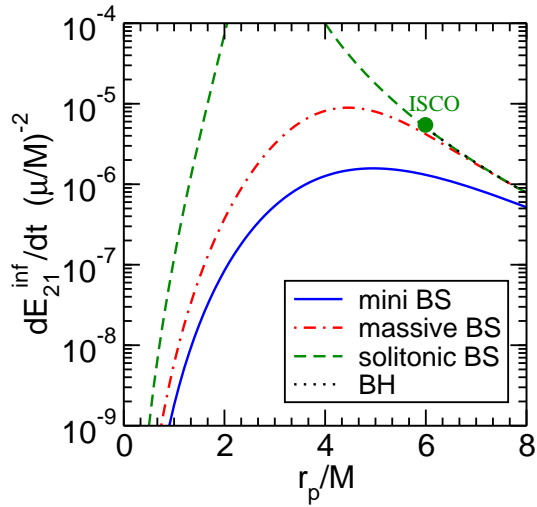}
\end{tabular}
\end{center}
\caption{\label{fig:axial_flux}
Dominant $l=2$, $m=1$ contribution to the axial gravitational flux emitted by a point-particle orbiting a BS for the stable BS configurations used in this work, compared to that of a Schwarzschild BH. The most compact configurations are closer to the BH case, and both solitonic configurations for $r>3M$ have basically the same values of the BH case.
}
\end{figure}

The axial sector is governed by a single equation which does not involve scalar perturbations. 
Due to the explicit form of the source term, the axial flux is vanishing for even values of $l+m$.
In Fig.~\ref{fig:axial_flux}, we show the dominant $l=2$, $m=1$ contribution of the axial flux for various BS models and compared to that of a Schwarzschild BH. 

\begin{figure*}
\begin{center}
\epsfig{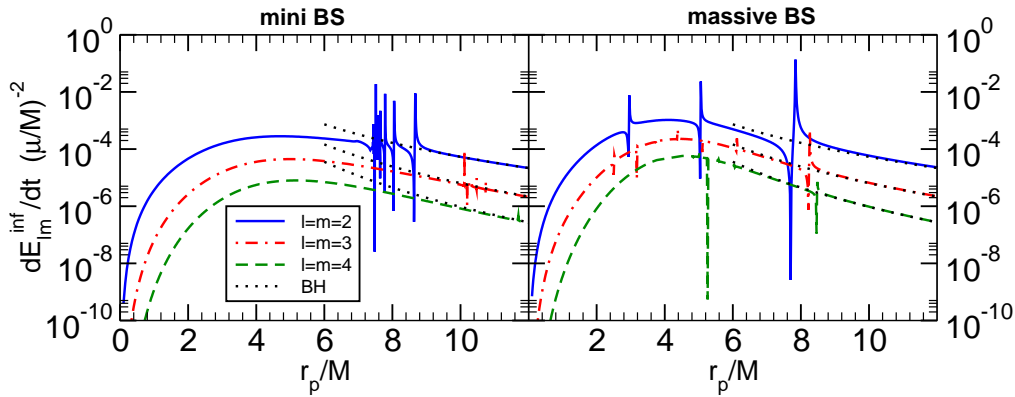}
\end{center}
\caption{\label{fig:polar_flux}
Main multipole contributions,  $l,m=2,3$ and $4$, for the mini and massive-BS configurations.}
\end{figure*}

The polar sector is described by an inhomogeneous system of coupled equations. A general method to solve this class of problems was presented in~\citep{Pani:2011xj} (see also~\citep{Molina:2010fb}). The contribution of the polar radiation to the total energy flux is nonvanishing only when $l+m$ is even and it is maximum when $l=m$. The dominant contributions is shown in Fig.~\ref{fig:polar_flux} as a function of the orbital distance for mini BSs and massive BSs. The axial and polar sector present similar features: at large distance the deviations from the BH case are basically indistinguishable, whereas for stable circular orbits inside the BS the energy flux quickly decreases as the orbit shrinks. In this limit, the orbital velocity is nonrelativistic and our results agree with a simple quadrupole formula~\citep{Maggiore:1900zz}.

However, the polar flux shows some sharp peaks which correspond to specific resonant frequencies. This interesting phenomenon occurs quite generically for small objects orbiting relativistic compact stars (see, e.g.~\citep{Pons:2001xs,Gualtieri:2001cm}) and it is related to the excitation of the quasinormal mode (QNM) frequencies. For a point-particle with orbital frequency $\Omega$, the resonance condition reads
\be
m \Omega =\sigma_R,
\ee
where $m$ is the azimuthal number and $\sigma_R$ is the real part of the QNM frequency. In other words, if the characteristic frequency of the BS matches (multiples of) the orbital frequency of the particle, sharp peaks appear in the emitted flux. This can be modeled in terms of a simple harmonic oscillator, where the orbiting particle acts as an external force and where the $\sigma_R$ is the proper frequency of the system. In this picture, the imaginary part of the frequency $\sigma_I$ is related to the damping of the oscillator and it is roughly proportional to the width of the resonance, while the quality factor $\sigma_R/\sigma_I$ is proportional to the square root of the resonance height~\citep{Pons:2001xs}. In agreement with this model, the resonances shown in Fig.~\ref{fig:polar_flux} correspond to the lowest damped QNMs, which are presented in Table~\ref{tab:BS_polar}. The QNM spectrum of a BS is rich and comprises several classes of modes~\citep{Yoshida:1994xi}. Here we have found a novel class of lowest damped modes, which can be excited during the inspiral due to their low frequency. A complete analysis of the QNM spectrum will appear in a separate work~\citep{elsewhere}.
\begin{table}[!ht]
 \begin{tabular}{c | c | c | c | c | c }
 Model   & $\sigma_R M$ & $-\sigma_I M$ &	$h/\nu^2$	& $\delta\phi_{\rm GW}$ [rads]\\
 \hline\hline
mini-BS I &  0.0757  & $3\times 10^{-5}$  	& $0.01$	& $6\times 10^3$\\
\hline \hline
massive-BS I &  0.0909  & $6\times 10^{-5}$ 	& $0.13$	& $9\times 10^4$\\
massive-BS I &  0.1616  & $5\times 10^{-6}$ 	& $0.02$ 	& $2\times 10^2$\\
massive-BS I &  0.2136  & $1\times 10^{-7}$ 	& $0.007$	& $4\times 10^{-1}$\\
\hline \hline
 \end{tabular}
 \caption{Polar quasi-bound modes $\sigma=\sigma_R+i\sigma_I$ of mini-BS and massive-BS configurations corresponding to the resonances shown in Fig.~\ref{fig:polar_flux} for $l=2$ and computed by a direct integration method~\citep{elsewhere}. We also present the height of the resonance normalized by the mass ratio, $h/\nu^2$, and the total GW dephasing $\delta\phi_{\rm GW}$ as computed in Eq.~\eqref{dephasing} for $T_{\rm obs}=1$yr and $M=10^5 M_\odot$.}\label{tab:BS_polar}
\end{table}

As shown in Fig.~\ref{fig:polar_flux}, the resonant frequencies may correspond to a stable circular orbit located \emph{outside} the BS radius (as for the rightmost resonance in the right panel of Fig.~\ref{fig:polar_flux}) or may correspond to stable circular orbits \emph{inside} the BS (as in the mini-BS case shown in the left panel of Fig.~\ref{fig:polar_flux}). While resonant circular orbits also occurs around perfect-fluid stars~\citep{Pons:2001xs} and other BH mimickers~\citep{Pani:2010em}, the existence of resonant geodesics inside the compact object is peculiar of BSs, due to the absence of a well-defined surface and due to the existence of stable circular orbits inside the star. Similar results as those shown in Fig.~\ref{fig:polar_flux} may be derived for other choices of the parameters and for higher values of~$l$.

The existence of these resonances is intriguing, because they appear to be a generic feature of compact objects supported solely by the self-gravity of a scalar field.
For BSs, this novel class of modes corresponds to the scalar perturbations being localized close to the BS radius and decaying exponentially at infinity, while the gravitational perturbations propagate to infinity as plane waves. These modes are usually named ``quasi-bound'' states~\citep{Dolan:2007mj,Rosa:2011my,Pani:2012bp} and they are supported by the mass of the scalar field.
In fact, any sufficiently compact object can support this class of modes in its interior. Constant density stars can support bound-state modes (i.e. modes with purely real frequency) for massive scalar perturbations with $l>0$~\citep{elsewhere}. In the case of a BS, these modes acquire a small imaginary part which is related to the coupling between scalar and gravitational perturbations: even if the scalar flux is zero for bound-state modes, part of the energy carried by the scalar field can be converted into gravitational energy that is then dissipated at infinity through GWs. This also explains qualitatively why the imaginary part of these modes is small (i.e. why the resonances are generically narrow), because the dissipation mechanism is not efficient. 

The overall structure of the resonances is fairly rich and it depends on the values of $l$, $m$ and on the specific models. 
However, in line with the case of a Schwarzschild BH, the modes have a hydrogenic-like spectrum. In that case, the location and width of the resonances can be computed analytically in the small mass limit~\citep{Detweiler:1980uk,Cardoso:2011xi}. For a Schwarzschild BH the quasi-bound modes $\sigma=\sigma_R+i\sigma_I$ read
\begin{eqnarray}
 \sigma_R&\approx&\mu\left(1-\frac{M^2\mu^2}{2 (n+l+1)}\right)\,,\nn\\
 \sigma_I&\approx&-\frac{4^{1-2 l}\pi ^2 (M\mu)^{4l+6}}{M(1+l+n)^{2 (2+l)}} \frac{ (2l+n+1)!}{ \Gamma\left[\frac{1}{2}+l\right]^2 \Gamma\left[\frac{3}{2}+l\right]^2 n!}\,,\nn
\end{eqnarray}
where $n\geq0$ is the overtone number. Therefore, as $\sigma$ approaches $\sigma_R$ there is a multitude of modes that can be excited and their separation in orbital frequency vanishes in the large $l$ or large $n$ limit. In the same limit the imaginary part (and hence the width of the resonances) of the modes decreases very rapidly, as shown by the last equation above. Our results are in qualitative agreement with this behavior, although in the BS models we considered the scalar field mass is not small, cf. Table~\ref{tableconf}. Indeed, in the opposite regime when $\mu M\ll1$ the imaginary part is exponentially suppressed, at least when the background spacetime is a BH~\citep{Zouros:1979iw}. We have found the same exponential behavior in the case of solitonic BSs  (which have $\mu M={\cal O}(10)$, cf. Table~\ref{tableconf}). This makes it extremely challenging to compute the polar flux for solitonic BSs and the corresponding flux would show extremely narrow resonances. This is the reason why we do not show the polar flux for solitonic BSs in Fig.~\eqref{fig:polar_flux}
However, the effective mass for a complex scalar field also depends on $\omega$ and, correspondingly, the resonance condition is shifted~\citep{elsewhere}. Our analysis generically shows that the resonant frequencies are of the order
\begin{equation}
 \Omega_{\rm res}=\frac{\mu\mp\omega}{m}\,,\label{Omegares}
\end{equation}
and the width of the resonances decreases quickly for large values of $m$ and for the overtones. It would be interesting to derive an analytical formula similar to the one above, for BSs. Remarkably, the resonant frequencies only depend on the parameters of the theory -- namely on the scalar field mass and on $\omega$ -- so that possible detection of resonant fluxes can be used to discriminate among different models and to tell a BS from a supermassive BH.

Let us now estimate the prospects of observing such effects. A generic framework to study the detectability of narrow resonances has been developed by~\citep{Yunes:2011aa}. While the orbiting body crosses the resonance, the emitted energy flux increases by orders of magnitude and the inspiral proceeds faster.
Therefore, the main observational consequence is a dephasing of the GW signal induced by the orbital acceleration at the resonance. 
Following~\citep{Yunes:2011aa}, we model the energy flux as a top-hat function\footnote{Note that, as shown in Fig.~\ref{fig:polar_flux} and consistently with a simple harmonic oscillator model~\citep{Pons:2001xs}, the resonant flux consists of a resonance and an antiresonance. Both can be modelled by a top-hat function as in Eq.~\eqref{tophat}, but they would have opposite sign. However, the height of the antiresonance is of the order of the Newtonian flux or less, whereas the height of the resonance is large by orders of magnitude. Therefore, we can safely neglect the contribution from the antiresonance and focus on a single top-hat function as in Eq.~\eqref{tophat}.}:
\begin{equation}
 \dot{E}=\dot{E}_0+h(t){\cal H}\left[\delta t_{\rm res}^2-(t-t_{\rm res})^2\right]\,,\label{tophat}
\end{equation}
where $\dot{E}_0$ is the flux in absence of the resonance which is well approximated by the quadrupole formula at large distance, $h(t)$ is a time-dependent height of the resonance occurring at $t=t_{\rm res}$, $\delta t_{\rm res}$ is its duration and ${\cal H}$ is the Heaviside function. In our case $\delta t_{\rm res}$ is much shorter than the orbital period, thus the leading-order formula for the dephasing induced by the resonance reads~\citep{Yunes:2011aa}
\begin{equation}
 |\delta\phi_{\rm GW}|\sim\frac{5 h}{16\nu^2}\frac{M\Delta\Omega_{\rm res}}{(M\Omega_{\rm res})^{10/3}}\frac{T_{\rm obs}}{M}\,, \label{dephasing}
\end{equation}
where $h=h(t_{\rm res})$, $\nu=\mu_p/M\ll1$ and $\Delta\Omega_{\rm res}$ is the width of the resonance in the frequency space. We estimate $\Delta\Omega_{\rm res}\approx 2 \sigma_I$ and $\Omega_{\rm res}\approx\sigma_R$, where $\sigma_R$ and $\sigma_I$ are the real and the imaginary part of the QNM frequency. Finally, we obtain
\begin{eqnarray}
 |\delta\phi_{\rm GW}|&\approx& 8.6\times 10^3  {\rm rads} \left[\frac{10^5 M_\odot}{M}\right]\l[\frac{T_{\rm obs}}{1{\rm yr}}\r]\nn\\
 &&\times\l[\frac{h/\nu^2}{10^{-2}}\r]\l[\frac{\sigma_I M}{10^{-5}}\r]\left[\frac{0.1}{\sigma_R M}\right]^{10/3}\,,
\end{eqnarray}
where we have rescaled all quantities by typical values as obtained for the peak of the resonance and for the QNM frequencies (cf. Table~\ref{tab:BS_polar}). Space-based observatories like eLISA/NGO will be sensitive to variations in the GW phase of the order of one radian~\citep{AmaroSeoane:2012km,AmaroSeoane:2012je}. The estimate above shows that the dephasing induced by the flux resonance can easily be larger by orders of magnitude. In fact, such large dephasing implies that a matched-filtering search for EMRIs that uses general relativistic templates would likely miss the signal or detect it but extracting completely wrong physical parameters. Depending on the parameters of the model, the height and the width of the resonance can be smaller and, for very narrow resonances, the dephasing will be negligible. Nonetheless, the bottom line of our analysis is that the quasi-circular inspiral of small compact objects around supermassive BSs will leave potentially detectable imprints that would be missed if the central object is \emph{a priori} assumed to be a BH.

Once the energy fluxes are computed, the evolution of an EMRI can be modeled using an adiabatic approximation and a Teukolsky evolution~\citep{2007PhRvD..76j4005S,Hughes:1999bq}. This procedure will include dissipative effects to all Post-Newtonian orders, but it is valid only at first order in the mass ratio. While such relativistic approach (together with including self-force effects~\citep{Poisson:2003nc,Barack:2009ux}) is crucial for generating accurate templates, no qualitatively new features will arise with respect to the case of inspiral around a massive BH. Our goal here is to point out new effects that can be used as a smoking gun for scalar-field configurations. One effect is the appearance of resonances discussed above. Another effect is the existence of stable circular orbits inside the BS and the possibility that the small compact object proceeds all the way down to the origin, without plunging. This was discussed in Section~\ref{sec:newtonian} within a much simpler --~although more general~-- Newtonian analysis. 

%

\section{Conclusions and outlook}\label{sec:conclusions}
According to general relativity, if self-interacting fundamental scalar fields exist in nature they may collapse to form compact self-gravitating configurations whose mass ranges from one to billions of solar masses, depending upon the scalar potential. Future GW detectors will be sensitive to the signal emitted by neutron stars and solar-mass BHs orbiting supermassive objects like those powering active galactic nuclei. In this paper, we have investigated several distinctive features of the inspiral around supermassive scalar-field configurations in the extreme mass-ratio regime. 
Rather than working on a case-by-case analysis, we focused on generic features that can leave a characteristic imprint on the gravitational waveform and which are fairly independent from the microphysics of DM particles. 

Working in a Newtonian approximation, we have discussed the inspiral in the interior of very generic DM configurations. If the small compact object interacts purely gravitationally with the scalar field, its motion will be described by quasi-elliptical orbits whose secular evolution is driven by DM accretion and by gravitational drag. These effects dominate the inspiral and are responsible for a peculiar GW signal. If accretion dominates over dynamical friction, the signal has a nearly constant amplitude and nearly constant frequency at late times. This is markedly different from the classical plunge which would occur if the central object were a BH and it might be used to discriminate the supermassive objects in galactic nuclei and to probe DM. We have shown that, already at Newtonian level, the waveforms in the Fourier space are strongly sensitive on the DM density profile.

As an aside application, our results may also be relevant to study the GW signal from the inspiral of putative primordial BHs~\citep{Hawking:1971ei} in the interior of neutron stars or to study the inspiral around Kerr BHs endowed with bosonic clouds~\citep{Arvanitaki:2010sy,Yoshino:2012kn,Pani:2012vp,Witek:2012tr}.

Secondly, the motion of the small compact object in the exterior of the supermassive configuration is driven by the emission of gravitational and scalar waves, which are coupled to each other. Due to this coupling, a baryonic test-particle in quasi-circular motion can resonantly excite the scalar QNMs of the central DM object. We have demonstrated this by considering some specific models of relativistic BSs in spherical symmetry. These resonances appear as sharp peaks in the energy flux emitted in GWs and would result in a faster inspiral, leading to a dephasing which can have observational consequences for future detectors. The resonant frequencies are largely insensitive to the details of the system, and they mainly depend only on the mass of the scalar particle. We have shown that the resonances correspond to the excitation of a novel class of QNMs, which have a much longer lifetime and should therefore dominate the late-time signal during the gravitational collapse and during the ringdown. We have discussed the QNM spectrum and the linear response of the system at fully relativistic level.

The plethora of DM candidates, modified gravitational theories and models for BH mimickers makes it mandatory to select generic features such as those described here. More rigorous case-by-case analysis might be performed if some of these signatures are eventually detected. Indeed, our work can be extended in several directions. A rigorous treatment of the accretion and drag effects in the interior beyond the Newtonian approximation is lacking. Relativistic effects might be important at the interface close to the radius, where velocities may be moderately relativistic and the motion could be supersonic. Special relativity effects were included in~\citep{Barausse:2007ph} and it would be interesting to extend our work by using those results and also by including Post-Newtonian corrections to the accretion-driven inspiral. On the other hand, the late-time inspiral close to the origin is intrinsically nonrelativistic and we expect our results to be accurate in that regime. Furthermore, a simple extension of our analysis is to include noncircular motion at fully relativistic level (see e.g.~\citep{Martel:2003jj}, where the same extension has been performed for the inspiral around a BH). 
Our results also show that a relativistic analysis as that performed by~\citep{Kesden:2004qx} should be extended to include the effects of accretion and dynamical friction.
Finally, we focused here on nonspinning solutions, but spin will certainly play a crucial role during both the external and the internal inspiral. Besides allowing for a much richer family of geodesics in its exterior, a spinning, compact horizonless object might be unstable under the so-called ``ergoregion instability''~\citep{Friedman:1978,Cardoso:2007az,Cardoso:2008kj}. This effect can be used to constrain the compactness-spin parameter space of compact horizonless objects and might also leave an imprint on the GW signal emitted during the inspiral. In addition, it would be interesting to extend our analysis of the accretion and gravitational drag effects to the case of spinning objects, where the angular momentum affects the relative velocity of the binary. This might be particularly important in a relativistic analysis of compact objects, in which frame-dragging effects have to be taken into account.


While this study was in its last stages, a related work appeared in the literature \citep{Eda:2013gg}; while the assumptions of both works are different, and our framework more general, the broad conclusions are both optimistic: gravitational waves can be an efficient tool to study DM.

\begin{acknowledgements}
We are indebted to Emanuele Berti, Kazunari Eda, Luis Lehner, Joseph Silk and specially to Enrico Barausse for useful correspondence.
C. M. and L. C. acknowledge CAPES and CNPq for partial financial support. L. C. is grateful to CENTRA-IST for kind hospitality.
P.P. acknowledges financial support provided by the European Community 
through the Intra-European Marie Curie contract aStronGR-2011-298297 
and the kind hospitality of Kinki University in Osaka.
V.C. acknowledges partial financial
support provided under the European Union's FP7 ERC Starting Grant ``The dynamics of black holes:
testing the limits of Einstein's theory'' grant agreement no. DyBHo--256667.
Research at Perimeter Institute is supported by the Government of Canada 
through Industry Canada and by the Province of Ontario through the Ministry
of Economic Development and Innovation.
This work was supported by the NRHEP 295189 FP7-PEOPLE-2011-IRSES Grant, and by FCT-Portugal through projects
PTDC/FIS/098025/2008, PTDC/FIS/098032/2008, PTDC/FIS/116625/2010,
CERN/FP/116341/2010 and CERN/FP/123593/2011.
Computations were performed on the ``Baltasar Sete-Sois'' cluster at IST,
the cane cluster in Poland through PRACE DECI-7 ``Black hole dynamics
in metric theories of gravity'',  
on Altamira in Cantabria through BSC grant AECT-2012-3-0012,
on Caesaraugusta in Zaragoza through BSC grants AECT-2012-2-0014 and AECT-2012-3-0011,
XSEDE clusters SDSC Trestles and NICS Kraken
through NSF Grant~No.~PHY-090003, Finis Terrae through Grant
CESGA-ICTS-234.
\end{acknowledgements}

\appendix

\section{Appendix A: Radiation-driven inspiral in the exterior}\label{app:Newtonian_ext}
Within a Newtonian approximation the inspiral in the exterior of the central object is simple and we follow~\citep{Maggiore:1900zz}. In absence of dissipation, the motion is elliptical with the mass $M$ located at the focus and the eccentricity $e$ and the semi-major axis $b$ are conserved quantities. At lowest order, the secular evolution of $e(t)$ and $b(t)$ driven by the GW emission can be modelled by the simple quadrupole formula. Clearly, this approach neglects any truly relativistic effect like the resonances discussed in the main text. On the other hand, the quadrupole formula provides a good approximation at large distances, when the orbital velocity is nonrelativistic. At first order in the mass ratio $\mu_p/M\ll1$, the equations for the orbital evolution are
\begin{eqnarray}
 \dot e &=& -\frac{304}{15}e\frac{\mu_p  M^2}{b^4\left(1-e^2\right)^{5/2}}\left(1+\frac{121}{304}e^2\right)\,, \\
 \dot b &=&  -\frac{64}{5}\frac{\mu_p  M^2}{b^3\left(1-e^2\right)^{7/2}}\left(1+\frac{73}{24}e^2+\frac{37}{96}e^4\right)\,, \label{dotb2}\\
 \dot \varphi &=&\frac{(1+e\cos\varphi)^2}{1-e^2}\sqrt{\frac{M}{b^3}} \,,  
\end{eqnarray}
where $\varphi$ is the angle in polar coordinates $(r,\varphi)$ and the motion occurs in the $\theta=\pi/2$ plane. The evolution is obtained solving the system above with some initial condition for $e$, $b$ and $\varphi$. The radius and angular momentum of the orbit read
\begin{equation}
 r(t)=\frac{b(1 - e^2)}{1 + e \cos\varphi}\,,\qquad
 L^2=\mu_p^2 M b(1-e^2)\,.
\end{equation}
Finally, the GW signal produced during the evolution is governed by the functions $h_+$ and $h_\times$, which can be written in terms of $r(t)$, $\varphi(t)$ and of the relative position of the distant observer (see, Eqs.~\eqref{hp} and \eqref{hm} in the main text). Note that the adiabatical evolution is valid provided radiation-reaction effects are small and the orbits evolve on timescales
much longer than a typical orbital period. This is indeed the case in the EMRI limit, at least when the orbital separation is large. Relativistic corrections to the quadrupolar formula are indeed important in the final stages of the exterior inspiral, close to the radius of compact stars.

\section{Appendix B: Motion of particles in a spring-like potential}\label{app:spring_elipse}
In this section we discuss the motion of a small particle inside a Newtonian, constant density star assuming the particle interacts only gravitationally with the star.
This configuration should be a good approximation during the latest stages of the inspiral inside compact scalar configurations, where the density profile is nearly constant.  
The motion of a body inside a constant density medium is described by a gravitational
potential of the form 
\be
V(r)=-\beta \mu_p+\gamma\mu_p\,r^2\,. \label{defV}
\ee
Here $\gamma=M/(2R^3),\,\beta=3M/(2R)$, where $M,R$ are the mass and the radius of the constant density object and
$\mu_p$ is the test particle's mass. In polar coordinates, the motion of the body is described by
\bea
\mu_p\,r^2\dot{\varphi}&=&L\,,\\
\frac{1}{2}\mu_p\, \dot{r}^2+\frac{L^2}{2\mu_p\,r^2}+V(r)&=&E\,,
\eea
where $L,E$ are the conserved angular momentum and energy parameter, respectively.
We can re-express the above as
\be
\left(\frac{1}{r^2}\frac{dr}{d\varphi}\right)^2=\frac{2\mu_p\,E}{L^2}-\frac{1}{r^2}-\frac{2\mu_p\,V(r)}{L^2}=\frac{2}{k}-\frac{1}{r^2}-\frac{2\mu_p^2\gamma r^2}{L^2}\,,
\ee
where $k^{-1}={\mu_p(E+\beta\mu_p)/L^2}$. Changing to $z=1/r^2-1/k$, we get
\be
\left(\frac{dz}{d\varphi}\right)^2=-4z^2+4B^2\,,
\ee
where
\be
B=\sqrt{\frac{1}{k^2}-\frac{2\mu_p^2\gamma}{L^2}}\,.
\ee
The solution to this equation is $z=B\cos2\varphi$. Thus, we get
\be
\frac{1}{r^2}=\frac{1}{k}\left(1+kB\cos2\varphi\right)\,.
\ee
In cartesian coordinates, this can be expressed as the standard equation for an ellipse:
\be
\frac{x^2}{a^2}+\frac{y^2}{b^2}=1\,,
\ee
where $a$ and $b$ are the semi-minor and semi-major axis and
\be
a^2=\frac{k}{1+kB}\,,\quad b^2=\frac{k}{1-kB}\,.
\ee
Thus, we get the interesting result that test masses follow ellipses which are centered at the origin and whose eccentricity reads
\be
e\equiv \sqrt{1-a^2/b^2}=\sqrt{2kB/(1+kB)}\,.
\ee
Recalling that $a^2=b^2(1-e^2)$, we can re-express the energy and angular momentum in terms of the eccentricity and semi-major axis $b$ as 
\bea
E&=&-\beta\mu_p+\gamma\mu_p b^2(2-e^2)\,,\label{E_ellipse}\\
L^2&=&2\mu_p^2\gamma(1-e^2)\,b^4\,.\label{L_ellipse}
\eea

Finally, using $\dot A=r^2\dot{\varphi}/2=L/(2\mu_p)$ and integrating over one orbit we get $A=LT/(2\mu_p)$, with $T$
the orbital period and $A$ the area of an ellipse. Using $A=\pi a b$, we find the analog of Kepler's third law:
\be
T^2=\frac{2\pi^2}{\gamma(1-e^2)}\,, \label{period}
\ee
and we obtain the interesting result that the orbital period is completely independent from semi-major axis,
as could be anticipated from the circular orbit case. Note that $T\propto \rho^{-1/2}$, where $\rho$ is the density of the medium.

\bibliographystyle{hapj}
\bibliography{emriboson}
\end{document}